\documentclass[epj,nopacs]{svjour}
\usepackage{latexsym}
\usepackage{graphics}
\usepackage{amssymb}

\newcommand{\be}{\begin{equation}}
\newcommand{\ee}{\end{equation}}
\newcommand{\nn}{\nonumber}
\newcommand{\bea}{\begin{eqnarray}}
\newcommand{\eea}{\end{eqnarray}}

\journalname{EPJ C}
\begin{document}
\setcounter{page}{1}
\title{
 \vspace{-23pt}
 {\rightline{\small LU 2003/004, LPNHE 2003-03}}
 New method for calculating helicity amplitudes\\
 of jet-like QED processes for high-energy colliders \\
 {\large II. Processes with lepton pair production}}
\titlerunning{New method for calculating helicity amplitudes. II}
\authorrunning{C.~Carimalo et al.}
\author{C.~Carimalo \inst{1}\thanks{carimalo@in2p3.fr} %
\and
A.~Schiller \inst{2,}\thanks{Arwed.Schiller@itp.uni-leipzig.de} %
\and
V.G.~Serbo \inst{3,}\thanks{serbo@math.nsc.ru}}
\institute{LPNHE, IN2P3-CNRS, Universit\'e Paris VI, F-75252 Paris, France
\and
Institut f\"ur Theoretische Physik and NTZ, Universit\"at Leipzig, D-04109
Leipzig, Germany
\and
Novosibirsk State University, Novosibirsk, 630090 Russia
}
\date{March 31, 2003
}

\abstract{
As continuation of our previous paper~\cite{CSS1} we further develop our
new method for calculating  helicity amplitudes of jet-like QED
processes described by tree diagrams, applying it to lepton pair
production. This method consists in replacing spinor
structures for real and weakly virtual intermediate leptons
by simple transition vertices. New vertices are introduced
for the pair production case, and previous bremsstrahlung vertices are
generalized to include virtual photons inside the considered jet.
We present a diagrammatic approach that allows to write down in an
efficient way the leading helicity amplitudes, at tree level. The
obtained compact amplitudes are particularly suitable for numerical
calculations in jet-like kinematics. Several examples
with up to four particles in a jet are discussed in detail.
}

\maketitle

\section{Introduction}
\label{intro}

In paper~\cite{CSS1} we have studied quantum electrodynamics (QED)
bremsstrahlung processes\footnote{Below we shall quote formulae
and figures from this paper by a double numbering, for example,
Eq.~(1.21) and Fig.~1.3 means Eq.~(21) and Fig.~3 from
Ref.~\cite{CSS1}. For a more complete list of references consult
our previous work.}
\be
  e(p_1)+  e^{\pm}(p_2) \to e(p_3)+ \gamma(k_1)+\ldots+ \gamma(k_n)+
  e^{\pm}(p_4)
  \label{1}
\ee
in the jet-like kinematics: $E_i\gg m$ and $m\lesssim |{\vec
p}_{i\perp}| \ll E_i$ including the helicities of all particles. In
particular, we have considered in detail the emission of one, two and three
photons along the direction of the first initial lepton described by the
block diagrams of
Figs.~1.2, 1.9 and 1.11, respectively. The corresponding amplitudes have the
form  
\be
  M_{fi}= \frac{s}{q^2} J_1 J_2
  \label{2}
\ee
where the impact factors $J_1$ for the different bremsstrahlung
processes in the first jet have been found in Sections 4, 5 and 6
of that paper, while the second (trivial) impact factor for
reaction (\ref{1}) is
\be
  J_2= \mp\,\sqrt{8\pi \alpha}\, \delta_{\lambda_2 \lambda_4} \,
  {\rm e}^{{\rm i} (\lambda_2 \varphi_2- \lambda_4 \varphi_4)}
  \,.
  \label{3}
\ee
The main idea of our approximation consists in  replacing the
numerators of the lepton propagators of small virtuality by
vertices which are matrices with respect to lepton helicities.
These vertices $V(p,k)$, $\widetilde V(p,k)$ and $V(p)$ are simple
analytical expressions given in Sect.~3.1 of \cite{CSS1} and they
represent the building blocks for the whole amplitude of the
bremsstrahlung processes.

In the present paper we consider processes with lepton pair
production. All nontrivial points can be demonstrated considering
the pair production processes of order $e^3$ and $e^4$.

To order $e^3$ there is one process for the photo-produc\-tion of a
single lepton pair (Fig.~1.3). The corresponding impact factor can
be easily obtained from the impact factor for the single
bremsstrahlung of Fig. 1.2  by a transition to the cross-channel.
Performing that transformation for the impact factor $J_1$, all
amplitudes for the processes of Figs.~1.4--1.6 are given as simple
combinations of the obtained impact factors.

Among the other processes of order $e^4$  mentioned in paper
\cite{CSS1}, only the bremsstrahlung pair production (Fig. 1.8)
can be considered as a new process, since diagrams of Figs.~1.7
and 1.10 represent amplitudes which are the cross-channel amplitudes for the
direct processes of Figs. 1.8 and 1.9, respectively.

Therefore, we have to calculate the impact factor for the
bremsstrahlung pair production of Fig. 1.8 and to find out the
substitution rules for energy fractions, transverse momenta and
polarizations of the particles under crossing. In the
calculation additional vertices will be introduced and the
bremsstrahlung vertices of our previous paper have to be
generalized including virtual photons with helicity zero.

\begin{figure}[!htb]
  \begin{center}
    \unitlength=2.00mm
    \begin{picture}(28.0,16.0)
      \put(12.00,12.55){\circle{4.80}}
      \put( 0.00, 0.00){\line(1,0){12.00}}
      \put( 0.00, 0.00){\vector(1,0){6.4}}
      \put( 0.00,10.90){\line(1,0){10.15}}
      \put( 0.00,10.90){\vector(1,0){6.4}}
      \put(12.00, 0.00){\line(0,1){2.0}}
      \put(12.00, 2.40){\line(0,1){2.0}}
      \put(12.00, 4.80){\line(0,1){2.0}}
      \put(12.00, 4.80){\vector(0,1){1.40}}
      \put(12.00, 7.20){\line(0,1){2.0}}
      \put(12.00, 9.60){\line(0,1){0.5}}
      \put(12.00, 0.00){\vector(1,0){6.6}}
      \put(12.00, 0.00){\line(1,0){12.0}}
      \put(14.40,14.20){\line(-1,0){0.55}}
      \put(16.80,14.20){\line(-1,0){2.0}}
      \put(19.20,14.20){\line(-1,0){2.0}}
      \put(17.20,14.20){\vector(1,0){1.40}}
      \put(21.60,14.20){\line(-1,0){2.0}}
      \put(24.00,14.20){\line(-1,0){2.0}}
    {\thicklines
      \put(14.40,12.00){\line(1,0){9.65}}
      \put(14.40,13.10){\line(1,0){9.65}}}
      \put(17.00,12.00){\vector(1,0){1.60}}
      \put(19.00,13.10){\vector(-1,0){1.50}}
      \put(14.00,10.90){\vector(1,0){4.6}}
      \put(13.85,10.90){\line(1,0){10.15}}
      \put(1.00,12.00){\makebox(0,0)[cc]{$e$}}
      \put(26.00,12.00){\makebox(0,0)[cc]{$_{\mu^-}$}}
      \put(26.00,13.10){\makebox(0,0)[cc]{$_{\mu^+}$}}
      \put(26.00,14.20){\makebox(0,0)[cc]{$_{\gamma}$}}
      \put(26.00,10.90){\makebox(0,0)[cc]{$_{e}$}}
    \end{picture}
  \end{center}
  \caption{The block diagram for the process  $e e \to e e \mu^+ \mu^-
  \gamma$}
  \label{fig:1}
\end{figure}
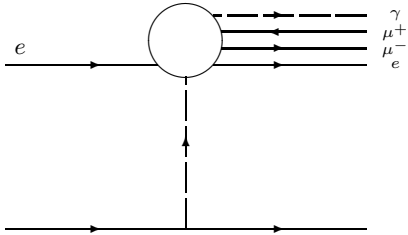

Those vertices as well as the rules for crossing are presented in
Sect.~\ref{sec:2}. In 
Sect.~\ref{sec:3} we derive the impact factor of Fig.~1.3 using the 
crossing relations which is the basis for pair production in 
$\gamma e$ and
$\gamma \gamma$ collisions. In the next Section we consider the bremsstrahlung
pair production of Fig.~1.8. Here for the first time intermediate virtual
photons have to be taken into account in the considered jet. This represents
the starting point for considering more complicated processes in the jet-like
kinematics. 
As an example illustrating the efficiency of the new method
we consider in Sect.~\ref{sec:5} its application to
the more complicated process of Fig.~\ref{fig:1} --- the collision
of an electron and a positron with production of a $\mu^+ \mu^-$
pair together with a photon inside the first jet. Our results are
summarized in the final section.

\section{Method for calculation of amplitudes with pair production}
\label{sec:2}

In addition to formulae presented in Sect.~2 of \cite{CSS1}, we
derive in this section some additional rules and formulae useful
for processes with lepton pair production.

\subsection{Vertices related to pair production}
\label{sec:2.1}

In this subsection we consider the new vertices appearing in the
processes described by Figs.~1.3, 1.5--1.8, 1.10 and Fig.~1. For
jet-like processes with lepton pair production we can repeat the
main points from~\cite{CSS1}. In particular, we present the
amplitude $M_{fi}$ in the factorized form (\ref{2}). Then, we
express the numerators of all spinor propagators $\hat{p}\pm m$
with $E>0$ for virtual $e^{\mp}$ via bispinors for real electrons
and positrons with 3-momentum ${\vec p}$ or $(-{\vec p})$:
\bea
  \hat p+m &=&  u_{\vec p}^{(\lambda)}
  \bar{u}_{\vec p}^{(\lambda)}+ {p^2
  -m^2\over 4 E^2} \, v_{-{\vec p}}^{(\lambda)} \bar{v}_{-{\vec
  p}}^{(\lambda)}\,,
  \label{4}\\
   \hat p-m &=&  v_{\vec p}^{(\lambda)} \bar{v}_{\vec p}^{(\lambda)}+ {p^2
  -m^2\over 4 E^2} \, u_{-{\vec p}}^{(\lambda)} \bar{u}_{-{\vec
  p}}^{(\lambda)}\,.
  \nn
\eea
This allows us to replace $\hat{p}\pm m$ by transition currents or
generalized vertices with real electrons and positrons. In
addition to the vertices $V(p,k)$, $\widetilde V(p,k)$ and $V(p)$
for bremsstrahlung given in (1.37)--(1.43), we have to consider
now new vertices for the bremsstrahlung of a virtual photon as
well as vertices for the  transition $\gamma(k) \to e^+(p_+) +
e^-(p_-)$ where $\gamma(k)$ is a real or virtual photon with
4-momentum $k$.

To illustrate this point, let us consider, for example, the
bremsstrahlung $\mu^- \mu^+$ production of Fig.~1.8
$$
  e^- + e^+ \to  e^- \mu^- \mu^+ + e^+ \,.
$$
\begin{figure}[!htb]
  \centering
  \unitlength=2.00mm
  \begin{picture}(25.00,16.00)
   \put(10.00,12.00){\circle{3.40}}
   \put( 0.50,10.90){\line(1,0){8.10}}
   \put( 0.50,10.90){\vector(1,0){5.15}}
   \put(10.00, 2.50){\line(0,1){2.0}}
   \put(10.00, 5.00){\line(0,1){2.0}}
   \put(10.00, 5.00){\vector(0,1){1.40}}
   \put(10.00, 7.50){\line(0,1){2.0}}
   \put(10.00,10.00){\line(0,1){0.2}}
   \put(15.50,13.10){\vector(4,1){2.9}}
   \put(20.50,11.85){\vector(-4,1){2.9}}
   \put(11.40,10.90){\line(1,0){ 9.1}}
   \put(11.40,10.90){\vector(1,0){5.00}}
   \put(11.40,13.10){\line(1,0){0.6}}
   \put(12.50,13.10){\line(1,0){2.0}}
   \put(15.00,13.10){\line(1,0){0.5}}
   \put(12.50,13.10){\vector(1,0){1.40}}
   \thicklines{
   \put(15.50,13.10){\line(4,1){5.0}}
   \put(15.50,13.10){\line(4,-1){5.0}}
   }
   \put(14.00,14.60){\makebox(0,0)[cc]{$k$}}
   \put(17.50, 8.90){\makebox(0,0)[cc]{$p_3$}}
   \put( 7.00, 8.90){\makebox(0,0)[cc]{$p_1$}}
   \put(11.50, 5.50){\makebox(0,0)[cc]{$q$}}
   \put(25.00, 8.50){\makebox(0,0)[cc]{$=$}}
   \put(22.50,14.50){\makebox(0,0)[cc]{$p_-$}}
   \put(22.50,12.00){\makebox(0,0)[cc]{$-p_+$}}
  \end{picture}
  \begin{picture}(44.00,16.00)
   \put(25.00, 8.50){\line(1,0){7.0}}
   \put(25.00, 8.50){\vector(1,0){2.90}}
   \put(29.66, 8.50){\vector(1,0){2.90}}
   \put(32.00, 8.50){\line(1,0){7.4}}
   \put(32.00, 8.50){\vector(1,0){5.30}}
   \put(29.66, 2.50){\line(0,1){2.00}}
   \put(29.66, 5.00){\line(0,1){2.00}}
   \put(29.66, 7.50){\line(0,1){1.00}}
   \put(29.66, 5.00){\vector(0,1){1.40}}
   \put(34.34,12.50){\line(0,1){0.60}}
   \put(34.34,10.00){\line(0,1){2.00}}
   \put(34.34, 8.50){\line(0,1){1.00}}
   \put(34.34,10.00){\vector(0,1){1.40}}
   \put(34.34,13.10){\vector(4,1){2.9}}
   \put(39.34,11.85){\vector(-4,1){2.9}}
   \put(32.84,11.20){\makebox(0,0)[cc]{$k$}}
   \put(28.16, 5.50){\makebox(0,0)[cc]{$q$}}
   \put(22.20, 8.50){\makebox(0,0)[cc]{$+$}}
   \put( 5.00, 8.50){\line(1,0){ 7.0}}
   \put( 5.00, 8.50){\vector(1,0){2.90}}
   \put( 9.66, 8.50){\vector(1,0){2.90}}
   \put(12.00, 8.50){\line(1,0){ 7.0}}
   \put(12.00, 8.50){\vector(1,0){5.20}}
   \put( 8.16,11.20){\makebox(0,0)[cc]{$k$}}
   \put(12.84, 5.50){\makebox(0,0)[cc]{$q$}}
   \put(14.34, 2.50){\line(0,1){2.00}}
   \put(14.34, 5.00){\line(0,1){2.00}}
   \put(14.34, 7.50){\line(0,1){1.00}}
   \put(14.34, 5.00){\vector(0,1){1.40}}
   \put( 9.66,12.50){\line(0,1){0.60}}
   \put( 9.66,10.00){\line(0,1){2.00}}
   \put( 9.66, 8.50){\line(0,1){1.00}}
   \put( 9.66,10.00){\vector(0,1){1.40}}
   \put(14.66,11.85){\vector(-4,1){2.9}}
   \put( 9.66,13.10){\vector(4,1){2.9}}
  {\thicklines
   \put( 9.66,13.10){\line(4,1){5.0}}
   \put( 9.66,13.10){\line(4,-1){5.0}}
   \put(34.34,13.10){\line(4,1){5.0}}
   \put(34.34,13.10){\line(4,-1){5.0}}}
  \end{picture}
  \caption{Feynman diagrams for the impact factor related to
           bremsstrahlung $\mu^+ \mu^-$ pair production
           }
  \label{fig:2}
\end{figure}
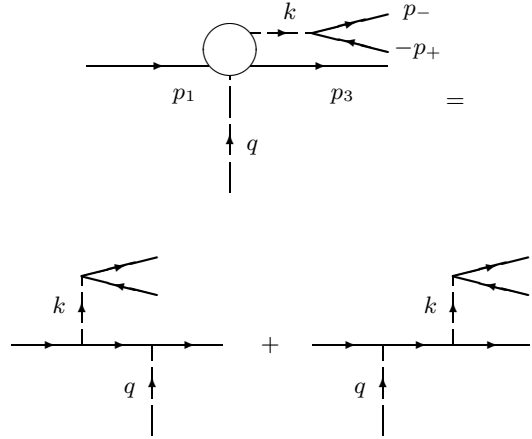
We present the impact factor of this process, described by Feynman
diagrams of Fig.~\ref{fig:2}, in the form
\be
  J_1 (e^-_{\lambda_1}+\gamma^* \to e^-_{\lambda_3}+
  \mu^+_{\lambda_+} +\mu^-_{\lambda_-})= {\sqrt{4\pi\alpha}\over
  k^2}\, A_\mu I^\mu\,,
  \label{5}
\ee
where $I^\mu$ is the current for the virtual transition $\gamma(k)
\to \mu^+(p_+) + \mu^-(p_-)$,
\be
  I^{\mu} =  \bar{u}_{{\vec p}_-}^{(\lambda_-)} \; \gamma^\mu \;
  {v}_{{\vec p}_+}^{(\lambda_+)}\,,
  \label{6}
\ee
and $A^\mu$ is the amplitude of the virtual Compton scattering,
\be
  A^\mu =  4\pi\alpha\left(\frac{N_1^\mu}{2p_1k-k^2} -
  \frac{N_3^\mu}{2p_3k+k^2}\right)\,,
  \label{7}
\ee
\bea
  N_1^\mu&=& \bar{u}_3 \, \hat{e}_q \,
  (\hat{p}_1-\hat k +m)\gamma^\mu u_1\,,
  \nn\\
  N_3^\mu&=& \bar u_3 \gamma^\mu (\hat{p}_3 +\hat k +m)
  \, \hat{e}_q \, u_1\,.
  \nn
\eea
The 4-vector
\be
  e_q \equiv {\sqrt{2} P_2\over s}
  \label{8}
\ee
can be considered as an ``effective polarization vector'' for the
$t$-channel virtual photon with momentum $q$.

Both 4-vectors $A^\mu$  and $I^\mu$ are orthogonal to the momentum
of the virtual photon $k$ due to gauge invariance:
$$
  Ak=Ik=0\,.
$$
Therefore, $A^\mu$ and $I^\mu$ have only three independent
components which can be chosen along the three polarization
4-vectors $e^{(\Lambda)}(k)$ for the virtual photon with
4-momentum $k$ and helicity $\Lambda =0,\,\pm 1$:
\bea
  {e}^{(\pm)}(k)&=&{e}_\perp^{(\pm)}-{k{e}_\perp^{(\pm)}\over
   kP_2}\,P_2\,,
  \label{9}\\
  {e}_\perp^{(\pm)}&=&\mp {1\over
  \sqrt{2}}\,(0,\,1,\,\pm {\rm i},\,0)\,,
  \nn\\
  {e}^{(0)}(k)&=& {\sqrt{k^2}\over kP_2}\left(P_2- {kP_2\over
  k^2}\,k\right)\,.
  \label{10}
\eea
These vectors obey the conditions:
\bea
  &&k\,e^{(\Lambda)}=0\,,\;\;
  e^{(\Lambda)*}e^{(\Lambda')}=-\delta_{\Lambda \Lambda'}\,,
  \nn\\
  &&\sum_{\Lambda=0,\,\pm 1} e_\mu^{(\Lambda)*}e_\nu^{(\Lambda)}=
  -g_{\mu \nu} +{k_\mu k_\nu\over k^2}\,.
  \label{11}
\eea
Due to gauge invariance, we can use $ {e}^{(0)}$ in the simpler form
\be
  {e}^{(0)}(k)= {\sqrt{k^2}\over kP_2}\,P_2\,.
  \label{12}
\ee
Note, that this vector is quite similar to the vector $e_q=\sqrt{2}
P_2/s$. The last equation in (\ref{11}) allows us to replace the
scalar product $AI$ by the sum over helicity states of the virtual
photon
\bea
  &&AI=-\sum_{\Lambda=0,\,\pm 1}A^{(\Lambda)} I^{(\Lambda)}\,,
  \label{13}
  \\
  &&A^{(\Lambda)}= A\, e^{(\Lambda)*}\,,  \quad
  I^{(\Lambda)}= I\, e^{(\Lambda)}\,.
 \nonumber
\eea
This relation is very convenient to analyse the structure of the
discussed impact factor.

The quantity $A^{(\Lambda)}$ has the same structure as the impact
factor for the single bremsstrahlung (1.61) and contains similar
vertices, but for photon helicities $\Lambda= 0,\,\pm 1$ and
photon virtuality $k^2\neq 0$. To obtain these vertices, we return
to (\ref{4}). The numerator of the spinor propagator $\hat p \pm m$ in
(\ref{4}) consists in two terms. The first term corresponds to the
simple replacements
\be
  \hat p+m \to  u_{\vec p}^{(\lambda)}\,\bar{u}_{\vec
  p}^{(\lambda)}\,,\;\; \hat p-m \to  v_{\vec p}^{(\lambda)}\,
  \bar{v}_{\vec p}^{(\lambda)}
  \label{14}
\ee
and leads to the vertex (1.43)
\bea
  V(p) &\equiv& {V}_{\lambda \lambda'} (p) =
  \bar{u}_{{\vec p}'}^{(\lambda')}\,  \hat {e}_q \,{u}_{\vec
  p}^{(\lambda)}=
  \bar{v}_{\vec p}^{(\lambda)}\,\hat {e}_q \,{v}_{{\vec p}'}^{(\lambda')}
  \nn\\
  &=& \sqrt{2}\, {E'\over E_1}\, \delta_{\lambda
  \lambda'}\, \Phi
  \label{15}
\eea
(with $E'=E$) and to the vertex
\bea
  V(p,\;k)&\equiv&  V_{\lambda \lambda'}^\Lambda(p,\;k) 
  \nn\\
  &=&
  \bar{u}_{{\vec p}'}^{(\lambda')}\; \hat {e}^{(\Lambda)\,*}(k)\;
  {u}_{\vec p}^{(\lambda)}=
  \bar{v}_{\vec p}^{(\lambda)}\; \hat {e}^{(\Lambda)\,*}(k)\;
  {v}_{{\vec p}'}^{(\lambda')}
  \nn\\
  &=&
  \Bigg\{ \bigg[ \delta_{\lambda \lambda'}\,
   2\, \left( {e}^{(\Lambda)\,*} p \right)\,
  \left(1- x\, \delta_{\Lambda,- 2\lambda} \right)
  \nn\\
  &+& \delta_{\lambda,-\lambda'}\,
  \delta_{\Lambda, 2\lambda}\, \sqrt{2}\, m x\bigg]
  \left( 1-\delta_{\Lambda,0} \right)
  \,
   \nn\\
    &+&  2  \sqrt{k^2}\, \frac{1-x}{x}\,
  \delta_{\lambda\lambda'}\,\delta_{\Lambda,0}  \Bigg\} \Phi\,.
  \label{16}
\eea
Here $x=\omega/E$ and the function $\Phi$ is the same as in
(1.41):
\be
  \Phi=\sqrt{\frac{E}{E'}}
  {\rm e}^{{\rm i} (\lambda' \varphi' - \lambda \varphi)}\,.
  \label{17}
 \ee
For helicity states $\Lambda = \pm 1$ the vertex (\ref{16})
coincides with that given in (1.39)\footnote{In the corresponding
formulae of \cite{CSS1} there are misprints: in (1.38) and in the
equation after (1.36) an additional ``minus'' sign has to be
inserted before the last equality; in (1.40) the factor $-\Lambda$
has to be eliminated; both signs ``minus'' have to be removed  in
(1.54); in (1.85), (1.87), (1.104), (1.105) each product of two
vertices ${\widetilde V}(p,\;k)$ should be taken with opposite
sign.}. It is convenient to present the vertices
(\ref{15})---(\ref{16}) by diagrams of Figs.~\ref{fig:3} and
\ref{fig:4}, respectively.
\begin{figure}[!htb]
  \begin{center}
  \unitlength=2.00mm
  \begin{picture}(44.00,9.00)
     \put( 0.00, 5.00){\line(1,0){10.00}}
     \put( 0.00, 5.00){\vector(1,0){5.4}}
     \put(10.00, 5.00){\line(1,0){10.00}}
     \put(10.00, 5.00){\vector(1,0){5.4}}
     \put(10.00, 3.80){\line(0,1){1.20}}
     \put(10.00, 1.90){\vector(0,1){1.00}}
     \put(10.00, 1.90){\line(0,1){1.20}}
     \put(10.00, 0.00){\line(0,1){1.20}}
     \put( 4.00, 6.50){\makebox(0,0)[cc]{$p$, $\lambda$}}
     \put(16.00, 6.50){\makebox(0,0)[cc]{$p'$, $\lambda'$}}
     \put(11.50, 2.00){\makebox(0,0)[cc]{$q$}}
     \put(22.00, 9.00){\makebox(0,0)[cc]{$V_{\lambda\lambda'}(p)$}}
     \put(24.00, 5.00){\line(1,0){10.00}}
     \put(44.00, 5.00){\vector(-1,0){5.4}}
     \put(34.00, 5.00){\line(1,0){10.00}}
     \put(34.00, 5.00){\vector(-1,0){5.4}}
     \put(34.00, 3.80){\line(0,1){1.20}}
     \put(34.00, 1.90){\vector(0,1){1.00}}
     \put(34.00, 1.90){\line(0,1){1.20}}
     \put(34.00, 0.00){\line(0,1){1.20}}
     \put(28.00, 6.50){\makebox(0,0)[cc]{$-p$, $\lambda$}}
     \put(40.00, 6.50){\makebox(0,0)[cc]{$-p'$, $\lambda'$}}
     \put(35.50, 2.00){\makebox(0,0)[cc]{$q$}}
  \end{picture}
  \end{center}
  \caption{Vertex $V(p)$ for initial
  electron (positron) with momentum $p$ and helicity $\lambda$,
  final electron (positron) with $p'$ and $\lambda'$ and
  $t$-channel virtual photon with momentum $q$
  and  ``effective polarization vector'' $e_q=\sqrt{2} P_2/s$
  }
  \label{fig:3}
\end{figure}
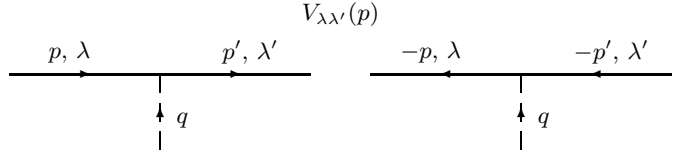
\begin{figure}[!htb]
  \begin{center}
  \unitlength=2.00mm
  \begin{picture}(44.00,8.00)
     \put( 0.00, 0.00){\line(1,0){10.00}}
     \put( 0.00, 0.00){\vector(1,0){5.4}}
     \put(10.00, 0.00){\line(1,0){10.00}}
     \put(10.00, 0.00){\vector(1,0){5.4}}
     \put(10.00, 0.00){\line(1,1){1.80}}
     \put(12.50, 2.50){\line(1,1){1.80}}
     \put(12.50, 2.50){\vector(1,1){1.20}}
     \put(15.00, 5.00){\line(1,1){1.80}}
     \put( 4.00, 1.50){\makebox(0,0)[cc]{$p$, $\lambda$}}
     \put(16.00, 1.50){\makebox(0,0)[cc]{$p'$, $\lambda'$}}
     \put(11.50, 5.00){\makebox(0,0)[cc]{$k$, $\Lambda$}}
     \put(24.00, 0.00){\line(1,0){10.00}}
     \put(44.00, 0.00){\vector(-1,0){5.4}}
     \put(34.00, 0.00){\line(1,0){10.00}}
     \put(34.00, 0.00){\vector(-1,0){5.4}}
     \put(34.00, 0.00){\line(1,1){1.80}}
     \put(36.50, 2.50){\line(1,1){1.80}}
     \put(36.50, 2.50){\vector(1,1){1.20}}
     \put(39.00, 5.00){\line(1,1){1.80}}
     \put(28.00, 1.50){\makebox(0,0)[cc]{$-p$, $\lambda$}}
     \put(40.00, 1.50){\makebox(0,0)[cc]{$-p'$, $\lambda'$}}
     \put(35.50, 5.00){\makebox(0,0)[cc]{$k$, $\Lambda$}}
     \put(23.00, 9.00){\makebox(0,0)[cc]
     {$V_{\lambda\lambda'}^\Lambda(p,\;k)$}}
  \end{picture}
  \end{center}
  \caption{Vertex $V(p,\,k)$  for initial
  electron (positron) with $p$ and $\lambda$,
  final electron (positron) with $p'$ and $\lambda'$ and
  final photon with $k$ and $\Lambda=0,\pm 1$
  }
  \label{fig:4}
\end{figure}
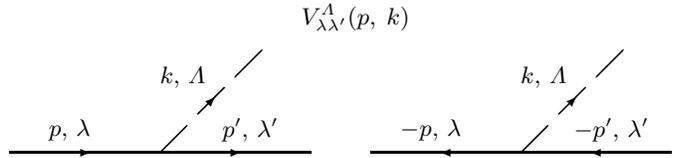

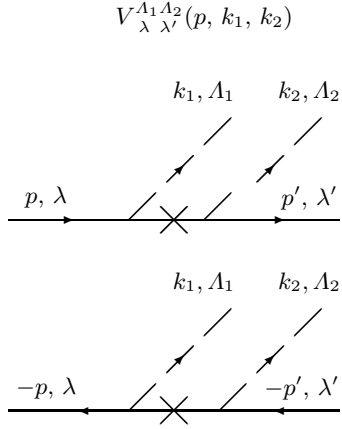
\begin{figure}[!htb]
  \begin{center}
  \unitlength=2.00mm
  \begin{picture}(25.0,15.0)
     \put(13.00, 13.50){\makebox(0,0)[cc]{
       $V_{\;\,\lambda \;\,\lambda'}^{\Lambda_1\Lambda_2}(p,\,k_1,\,k_2)$
      }}
     \put( 0.00, 0.00){\line(1,0){8.00}}
     \put( 0.00, 0.00){\vector(1,0){4.4}}
     \put( 8.00, 0.00){\line(1,0){6.00}}
     \put(10.10,-0.90){\line(1,1){1.80}}
     \put(10.10, 0.90){\line(1,-1){1.80}}
     \put(14.00, 0.00){\line(1,0){8.00}}
     \put(14.00, 0.00){\vector(1,0){4.4}}
     \put( 8.00, 0.00){\line(1,1){1.80}}
     \put(10.50, 2.50){\line(1,1){1.80}}
     \put(10.50, 2.50){\vector(1,1){1.20}}
     \put(13.00, 5.00){\line(1,1){1.80}}
     \put(13.00, 0.00){\line(1,1){1.80}}
     \put(16.50, 2.50){\line(1,1){1.80}}
     \put(16.50, 2.50){\vector(1,1){1.20}}
     \put(19.00, 5.00){\line(1,1){1.80}}
     \put( 2.50, 1.50){\makebox(0,0)[cc]{$p$, $\lambda$}}
     \put(20.00, 1.50){\makebox(0,0)[cc]{$p'$, $\lambda'$}}
     \put(13.00, 8.50){\makebox(0,0)[cc]{$k_1, \Lambda_1$}}
     \put(20.00, 8.50){\makebox(0,0)[cc]{$k_2, \Lambda_2$}}
  \end{picture}
  \begin{picture}(25.0,12.50) 
     \put( 0.00, 0.00){\line(1,0){8.00}}
     \put( 9.00, 0.00){\vector(-1,0){4.4}}
     \put( 8.00, 0.00){\line(1,0){6.00}}
     \put(10.10,-0.90){\line(1,1){1.80}}
     \put(10.10, 0.90){\line(1,-1){1.80}}
     \put(14.00, 0.00){\line(1,0){8.00}}
     \put(22.00, 0.00){\vector(-1,0){4.4}}
     \put( 8.00, 0.00){\line(1,1){1.80}}
     \put(10.50, 2.50){\line(1,1){1.80}}
     \put(10.50, 2.50){\vector(1,1){1.20}}
     \put(13.00, 5.00){\line(1,1){1.80}}
     \put(14.00, 0.00){\line(1,1){1.80}}
     \put(16.50, 2.50){\line(1,1){1.80}}
     \put(16.50, 2.50){\vector(1,1){1.20}}
     \put(19.00, 5.00){\line(1,1){1.80}}
     \put( 2.50, 1.50){\makebox(0,0)[cc]{$-p$, $\lambda$}}
     \put(19.50, 1.50){\makebox(0,0)[cc]{$-p'$, $\lambda'$}}
     \put(13.00, 8.50){\makebox(0,0)[cc]{$k_1, \Lambda_1$}}
     \put(20.00, 8.50){\makebox(0,0)[cc]{$k_2, \Lambda_2$}}
  \end{picture}
  \end{center}
  \caption{
  Vertex $V(p,k_1,k_2)$
  for initial and final electrons
  (positrons) with $p$, $\lambda$ and $p'$,
  $\lambda'$ and emission of two photons with $k_{1}$,
  $\Lambda_{1}$ and $k_{2}$, $\Lambda_{2}$ connected by
  an crossed intermediate
  lepton line
  }
  \label{fig:5}
\end{figure}

The second term in $\hat p \pm m$ of (\ref{4}) corresponds to the more
complicated replacements [cf. (1.32)]
\bea
  \hat p+m &\to&  {p^2-m^2\over 4 E^2} \, v_{-{\vec p}}^{(\lambda)}
  \bar{v}_{-{\vec p}}^{(\lambda)} \approx {p^2-m^2\over 4 EE_2}\,
  {\hat P}_2\,,
  \nn\\
  \hat p-m &\to& {p^2-m^2\over 4 E^2} \, u_{-{\vec p}}^{(\lambda)}
  \bar{u}_{-{\vec p}}^{(\lambda)} \approx {p^2-m^2\over 4 EE_2}\,
  {\hat P}_2\,.
  \label{18}
\eea
Since these expressions contain a factor proportional to the
denominator of the spinor propagator,
that denominator is cancelled and a new vertex with four external lines
can be introduced 
(similar to a vertex with four external particles in
scalar QED). Graphically we denote this vertex by the diagrams of
Fig.~\ref{fig:5} in which the crossed lepton line represents the
contracted line corresponding to replacement (\ref{18}).
Using this newly defined vertex, we can avoid from now on
vertices $\widetilde{V}$
used in \cite{CSS1}.

The vertex of Fig.~\ref{fig:5} is given by
\bea
  &&V(p,k_1,k_2) \equiv V_{\;\,\lambda
  \;\,\lambda'}^{\Lambda_1\Lambda_2}(p,k_1,k_2)
  \nn\\
  &=&
  {1\over 4(E-\omega_1)E_2}\, \bar{u}_{{\vec p}'}^{(\lambda')} \;
  \hat {e}^{(\Lambda_2)*}(k_2)\; \hat{P}_2 \;
  \hat{e}^{(\Lambda_1)*}(k_1)\; {u}_{{\vec p}}^{(\lambda)}
  \nn\\
  &=&
  {1\over 4(E-\omega_1)E_2}\, \bar{v}_{{\vec p}}^{(\lambda)} \; 
  \hat {e}^{(\Lambda_1)*}(k_1)\; \hat{P}_2 \;
  \hat{e}^{(\Lambda_2)*}(k_2)\; {v}_{{\vec p}'}^{(\lambda')}
  \nn\\
  &=&
  - 2{E'\over E-\omega_1}\,
  \delta_{\lambda \lambda'}\delta_{\Lambda_1, 2\lambda}
  \,\delta_{\Lambda_1,-\Lambda_2}
  \Phi.
  \label{19}
\eea
This equation can be easily proved if we notice that for the used
polarization vectors (\ref{9}) and (\ref{12}) we have
\bea
  \hat {e}^{(\Lambda_2)*}(k_2) & \hat{P}_2& 
  \hat{e}^{(\Lambda_1)*}(k_1)
  \nn\\
  &=& \hat {e}^{(\Lambda_2)*}_\perp\; \hat{P}_2 \;
  \hat{e}^{(\Lambda_1)*}_\perp\; \left(1-\delta_{\Lambda_1,0}\right)\,
  \delta_{\Lambda_1,-\Lambda_2}
  \nn\\
  &=&- \hat{P}_2 \;\left(1+\Lambda_1\, \Sigma_z\right)\,
  \delta_{\Lambda_1, \pm 1}\,\delta_{\Lambda_2, \mp 1}\,,
  \label{20}
\eea
where the matrix $\Sigma_z$ is diagonal in the standard as well as
in the spinor representation
$$
  \Sigma_z=\left(
  \begin{array}{cc}
    \sigma_z & 0 \\
    0& \sigma_z  
  \end{array}
  \right)\,,
$$
$\sigma_z$ is the Pauli matrix.

In the spirit of paper\cite{CSS1}, we can introduce the following
additional diagrammatic rules:
\begin{enumerate}
  \item
  A crossed lepton line cannot start or end at a vertex with the
  $t$-channel virtual photon of momentum $q$ or a virtual photon in
  the considered jet with helicity state $\Lambda=0$ (other helicity
  states for virtual photons of the jet are allowed).
  \item
  Two crossed lines should be separated at least by one uncrossed
  line. In other words, vertices with more than four external lines
  are forbidden.
\end{enumerate}
Those rules are due to the fact that a crossed line corresponds to
replacements (\ref{18}). Therefore, rule (1) is due to the
equations $\hat{e}_q\,\hat{P}_2= \hat{P}_2\,\hat{e}_q=0$ with
$e_q=\sqrt{2} P_2/s$ and $\hat{e}^{(0)}\, \hat{P}_2=\hat{P}_2\,
\hat{e}^{(0)}=0$. Rule (2) is a consequence of the  equation
$\hat{P}_2\, \hat{e}^{(\Lambda)}(k)\, \hat{P}_2=0$.

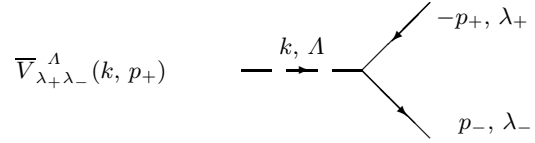
\begin{figure}[!htb]
  \begin{center}
  \unitlength=2.00mm
  \begin{picture}(25.00,10.00)
     \put(10.00, 5.00){\line(1,0){2.00}}
     \put(13.00, 5.00){\line(1,0){2.00}}
     \put(13.00, 5.00){\vector(1,0){1.50}}
     \put(16.00, 5.00){\line(1,0){2.00}}
     \put(18.00, 5.00){\line(1,1){4.50}}
     \put(22.50, 9.50){\vector(-1,-1){2.50}}
     \put(18.00, 5.00){\line(1,-1){4.50}}
     \put(18.00, 5.00){\vector(1,-1){3.0}}
     \put(26.00, 8.50){\makebox(0,0)[cc]{$-p_+$, $\lambda_+$}}
     \put(26.90, 1.50){\makebox(0,0)[cc]{$p_-$, $\lambda_-$}}
     \put(14.00, 6.50){\makebox(0,0)[cc]{$k$, $\Lambda$}}
     \put( 0.00, 5.00){\makebox(0,0)[cc]{
       $\overline{V}_{\lambda_+\lambda_-}^{\;\;\Lambda} (k,\,p_+)$
       }}
 \end{picture}
  \end{center}
  \caption{Vertex $\overline{V}(k,\,p_+)$
   for initial photon ($k$, $\Lambda$)
   and final lepton ($p_-$, $\lambda_-$) and
   antilepton ($p_+$, $\lambda_+$)}
   \label{fig:6}
\end{figure}

Now we consider the current $I^{(\Lambda)}$ (\ref{6}) which is
a new vertex corresponding to the transitions 
\\
$\gamma(k) \to  \mu^+(p_+) + \mu^-(p_-)$ or
$\gamma(k) \to e^+(p_+) + e^-(p_-) $
with $k=p_+ + p_-$ (Fig.~\ref{fig:6}):
\be
  \overline{V} (k,\,p_+)\equiv I^{(\Lambda)}=
  \overline{V}_{\lambda_+ \lambda_-}^{\,\Lambda}(k,\,p_+)=
  \bar{u}_{{\vec p}_-}^{(\lambda_-)} \; \hat{e}^{(\Lambda)}(k)\;
  {v}_{{\vec p}_+}^{(\lambda_+)}\,.
  \label{21}
\ee

Let us compare this vertex  with  $V(p,\,k)$ from (\ref{16}).
Taking into account that
$$
  {e}^{(\Lambda)}_\perp = -{e}_\perp^{(-\Lambda)\,*}\,,\;\;
  {e}^{(0)}(k) =-{e}^{(0)}(-k)\,,
$$
we find
\be
  {e}^{(\Lambda)}(k)= -{e}^{(-\Lambda)\,*}(-k)\,.
  \label{22}
\ee
In addition, it is known (see Appendix in \cite{CSS1}) that the
bispinor $v^{(\lambda_+)}_{{\vec p}_+}$ can be obtained from the
bispinor $u^{(\lambda)}_{\vec p}$:
\be
  v^{(\lambda_+)}_{{\vec p}_+}=u^{(\lambda)}_{\vec p}(E \to -E_+,\,
  \lambda \to -\lambda_+,\, \theta \to \theta_+,\, \varphi \to
  \varphi_+)
  \label{23}
\ee
with the convention
$\sqrt{-E_+} = +{\rm i}\sqrt{E_+}$. Due to these connections
between spinors and polarization vectors, we find the useful
relation
\be
  \overline{V}_{\lambda_+ \lambda_-}^{\,\Lambda}(k,\,p_+)=-
  {V}_{-\lambda_+ \lambda_-}^{-\Lambda}(-p_+,\,-k)
  \label{24}
\ee
from which it immediately follows that
\bea
  &&\overline{V}^{\;\Lambda}_{\lambda_+ \lambda_-}(k,\,p_+)
  \nn\\
  &&=
  {\rm i}\,
  \Bigg\{ {1\over x_+}\, \bigg[\delta_{\lambda_+, -\lambda_-}\,
  2\, \left( {e}^{(\Lambda)} p_+\right)\, \left(\delta_{\Lambda,-
  2\lambda_+} - x_+ \right)\Biggr.\Bigr.
  \nn\\
  &&- \Biggl.\Bigl. \delta_{\lambda_+
  \lambda_-}\,\delta_{\Lambda, 2\lambda_+}\, \sqrt{2}\, m \,\bigg]\,
  \left(1-\delta_{\Lambda,0}\right) 
  \label{25}\\
  &&+
    2\,\sqrt{k^2}\, x_-\,
  \delta_{\lambda_+,- \lambda_-}\,\delta_{\Lambda,0}
  \Bigg\}\,
   \overline\Phi\,,
  \nn
\eea
where
\be
  x_{\pm}={E_{\pm}\over \omega}\,,\;\;\;
  \overline\Phi = \sqrt{E_+\over E_-}\, {\rm
  e}^{{\rm i} (\lambda_+ \varphi_+ + \lambda_- \varphi_-)}\,.
  \label{26}
\ee

It is useful to recall that for the polarization vectors we
have
\bea
  {e}^{(\pm)}p_+ &=& {e}^{(\pm)}_\perp\, \left(p_{+\perp} -
  x_+{k_\perp} \right)
  \nn \\
  &=& -{e}^{(\pm)}p_- = -{e}^{(\pm)}_\perp\,
  \left(p_{-\perp} - x_-{k_\perp} \right)\,.
  \label{27}
\eea
Using (\ref{27}) and the equalities $x_++ x_-=1$, $k_\perp
=p_{+\perp}+p_{-\perp}$, we can rewrite (\ref{25}) in the form
\bea
  &&  \overline{V}^{\;\Lambda}_{\lambda_+ \lambda_-}(k,\,p_+)
  =
  {\rm i} \, \sqrt{E_+ E_-}
  \nn\\
  &&\times
  \Bigg\{
  \Bigg[2 \delta_{\lambda_+, -\lambda_-}   {e}^{(\Lambda)}_\perp
  \left( \frac{p_{+\perp}}{E_+} \delta_{\Lambda,-2\lambda_+}+
         \frac{p_{-\perp}}{E_-} \delta_{\Lambda,-2\lambda_-}-
         \frac{k_\perp}{\omega} \right)
   \nn\\
  &&- \delta_{\lambda_+\lambda_-}\,
    \delta_{\Lambda, 2\lambda_+}\,
    \sqrt{2}\,{m\omega\over E_+ E_-}
    \,\Bigg]\,\:
    \left(1-\delta_{\Lambda,0}\right)
  \nn\\
  &&  +2{\sqrt{k^2}\over \omega}\,\delta_{\lambda_+,-\lambda_-}\,
    \delta_{\Lambda,0}\,\Bigg\}\,
    {\rm e}^{{\rm i} (\lambda_+ \varphi_+ + \lambda_- \varphi_-)},
  \label{28}
\eea
which clearly exhibits the symmetry under lepton ex\-change $e^+
\leftrightarrow e^-$:
\be
  \overline{V}_{\lambda_+\, \lambda_-}^{\;\Lambda}(k,\;p_+)=
  \overline{V}_{\:\lambda_-\, \lambda_+}^{\;\Lambda}(k,\;p_-)\,.
  \label{29}\ee

\begin{figure}[!htb]
  \begin{center}
  \unitlength=2.00mm
  \begin{picture}(25.00,14.50)
     \put(12.00, 12.00){\makebox(0,0)[cc]
     {$\overline{V}_{\lambda_+ \lambda_-}^{\;\Lambda\,\Lambda'}
     (k,p_+,k')$}}
     \put( 0.00, 0.00){\line(1,0){2.20}}
     \put( 2.90, 0.00){\line(1,0){2.20}}
     \put( 2.90, 0.00){\vector(1,0){1.50}}
     \put( 5.80, 0.00){\line(1,0){2.20}}
     \put( 8.00, 0.00){\line(1,0){6.00}}
     \put(14.00, 0.00){\line(1,0){8.00}}
     \put(14.00, 0.00){\vector(1,0){4.4}}
     \put( 10.10,-0.90){\line(1,1){1.80}}
     \put( 10.10, 0.90){\line(1,-1){1.80}}
     \put( 8.00, 0.00){\line(1,1){6.80}}
     \put(14.80, 6.80){\vector(-1,-1){3.80}}
     \put(14.00, 0.00){\line(1,1){1.80}}
     \put(16.50, 2.50){\line(1,1){1.80}}
     \put(16.50, 2.50){\vector(1,1){1.20}}
     \put(19.00, 5.00){\line(1,1){1.80}}
     \put( 2.50, 1.50){\makebox(0,0)[cc]{$k$, $\Lambda$}}
     \put(20.00, 1.50){\makebox(0,0)[cc]{$p_-$, $\lambda_-$}}
     \put(14.00, 8.50){\makebox(0,0)[cc]{$-p_+$, $\lambda_+$}}
     \put(21.00, 8.50){\makebox(0,0)[cc]{$k'$, $\Lambda'$}}
  \end{picture}
  \begin{picture}(25.00,17.00)
     \put(12.00,12.00){\makebox(0,0)[cc]
     {$\overline{V}_{\lambda_- \lambda_+}^{\;\Lambda\,\Lambda'}
     (k,p_-,k')$}}
     \put( 0.00, 0.00){\line(1,0){2.20}}
     \put( 2.90, 0.00){\line(1,0){2.20}}
     \put( 2.90, 0.00){\vector(1,0){1.50}}
     \put( 5.80, 0.00){\line(1,0){2.20}}
     \put( 8.00, 0.00){\line(1,0){6.00}}
     \put(14.00, 0.00){\line(1,0){8.00}}
     \put(22.00, 0.00){\vector(-1,0){4.0}}
     \put(10.10,-0.90){\line(1,1){1.80}}
     \put(10.10, 0.90){\line(1,-1){1.80}}
     \put( 8.00, 0.00){\line(1,1){6.80}}
     \put( 8.00, 0.00){\vector(1,1){3.80}}
     \put(14.00, 0.00){\line(1,1){1.80}}
     \put(16.50, 2.50){\line(1,1){1.80}}
     \put(16.50, 2.50){\vector(1,1){1.20}}
     \put(19.00, 5.00){\line(1,1){1.80}}
     \put( 2.50, 1.50){\makebox(0,0)[cc]{$k$, $\Lambda$}}
     \put(20.00, 1.50){\makebox(0,0)[cc]{$-p_+$, $\lambda_+$}}
     \put(14.00, 8.50){\makebox(0,0)[cc]{$p_-$, $\lambda_-$}}
     \put(21.00, 8.50){\makebox(0,0)[cc]{$k'$, $\Lambda'$}}
  \end{picture}
  \end{center}
  \caption{
  Vertices $\overline{V}(k,\,p_+,k')$ and $\overline{V}(k,\,p_-,k')$
  for the transition of initial photon ($k$, $\Lambda$) to final
  electron (or positron) with emission of positron (or electron)
  and photon ($k'$, $\Lambda'$) connected by an crossed
  intermediate lepton line
  }
  \label{fig:7}
\end{figure}
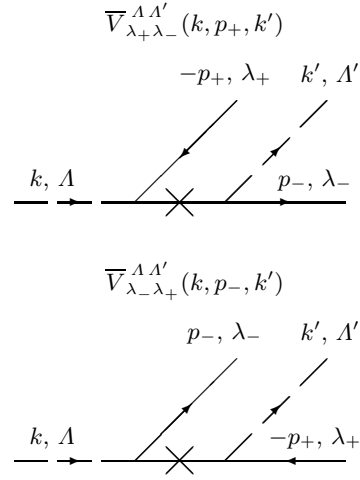

Analogously to (\ref{19}), it is convenient to introduce vertices
with four external lines obtained from diagrams of
Fig.~\ref{fig:5} by interchanging one of the outgoing photons with
the initial lepton or antilepton. For example, performing the replacements
$k_1 \to -k_1$, $\Lambda_1 \to -\Lambda_1 $, $p \to -p$, $\lambda
\to -\lambda$ we get
\bea
  && \overline{V}(k_1,\,p,\,k_2) \equiv \overline{V}_{\;\;\lambda\;\,
  \lambda'}^{\,\Lambda_1\Lambda_2}(k_1,\,p,\,k_2)
  \nn\\
  &&=
  {1\over 4(\omega_1-E)E_2}\, \bar{u}_{{\vec p}'}^{(\lambda')} \;
  \hat {e}^{(\Lambda_2)*}(k_2)\; \hat{P}_2 \;
  \hat{e}^{(\Lambda_1)}(k_1)\; {v}_{{\vec p}}^{(\lambda)}
  \nn\\
  &&=
  {1\over 4(\omega_1-E)E_2}\, \bar{u}_{{\vec p}}^{(\lambda)} \; \hat
  {e}^{(\Lambda_1)}(k_1)\; \hat{P}_2 \;
  \hat{e}^{(\Lambda_2)*}(k_2)\; {v}_{{\vec p}'}^{(\lambda')}
  \nn \\
  &=& 2{\rm i}{E'\over \omega_1-E}\,
  \delta_{\lambda, -\lambda'}\,\delta_{\Lambda_1, 2\lambda}
  \,\delta_{\Lambda_1,\Lambda_2}\, \sqrt{E\over E'} \,{\rm
  e}^{{\rm i} (\lambda \varphi + \lambda' \varphi')}\,.
  \label{30}
\eea
The last identity in this equation can be easily obtained using
a relation analogous to (\ref{24}):
\be
  \overline{V}_{\;\;\lambda\;\,\lambda'}^{\,\Lambda_1\Lambda_2}
  (k_1,\,p,\,k_2)=-{V}_{\;-\lambda\;\,\lambda'}^{-\Lambda_1\Lambda_2}
  (-p,\,-k_1,\,k_2)\,.
  \label{30a}
\ee
Furthermore, we can specify equation (\ref{30}) for the two diagrams
of Fig.~\ref{fig:7} which represent the two possible crossings of the
diagrams of Fig.~\ref{fig:5}:
either
\bea
  &&\overline{V}_{\lambda_+ \lambda_-}^{\;\Lambda
  \Lambda'}(k,\;p_+,\;k')
  \label{31}\\
  && = 2{\rm i}\, \frac{\sqrt{E_+E_-}}{\omega-E_+}\,
  \delta_{\lambda_+, -\lambda_-}\,\delta_{\Lambda, 2\lambda_+}
  \,\delta_{\Lambda \Lambda'}\, 
  {\rm e}^{{\rm i} (\lambda_+ \varphi_+ + \lambda_- \varphi_-)} 
  \nn
\eea
or
\bea
  &&\overline{V}_{\lambda_- \lambda_+}^{\;\Lambda
  \Lambda'}(k,\;p_-,\;k')
  \label{32}\\
  &&= 2{\rm i}\, \frac{\sqrt{E_+E_-}}{\omega-E_-} \,
  \delta_{\lambda_+, -\lambda_-}\,\delta_{\Lambda, 2\lambda_-}
  \,\delta_{\Lambda \Lambda'}\,
  {\rm e}^{{\rm i} (\lambda_+ \varphi_+ + \lambda_- \varphi_-)}\,.
  \nn
\eea

The further calculation of the discussed impact factor (\ref{5})
will be performed in Sect.~\ref{sec:4}.

\subsection{Some properties of the vertex $\overline{V}(k,\,p_+)$}
\label{sec:2.2}

For reference reasons it is useful to present formulae (\ref{25})
for some particular cases, omitting the factors $\overline{\Phi}$.
In the case of the helicity conserved transitions, $\lambda_+ =
-\lambda_-$, these vertices are of the form:
\bea
  \overline{V}(k,p_+)& =&-2{\rm i} \; e p_+
  \;\;\;\;\;\;\;\;\;\;\;\;\;\;\,
  \mbox{ for } \;
  \Lambda =2\lambda_+ =-2 \lambda_-\,,
  \nn\\
   \overline{V}(k,p_+) &=&{2{\rm i}\over x_+}\, (1-x_+)\,e p_+ \; 
  \mbox{ for } \; \Lambda
  =-2\lambda_+ =2 \lambda_-\,,
  \nn\\
  \overline{V}(k,p_+) &=& 2{\rm i}\sqrt{k^2}\, x_-
  \;\;\;\;\;\;\;\;\;\;\;\;\,
   \mbox{ for } \; \Lambda  =0\,.
   \label{34}
\eea
In the case of the helicity non-conserved transitions, $\lambda_+
= \lambda_-$, we have:
\bea
  \overline{V}_{++}^{\;+}(k,p_+) &=&
  \overline{V}_{--}^{\;-}(k,p_+) =- {\sqrt{2}\,{\rm i}\over x_+}\, m\,,
  \nn\\
  \overline{V}_{++}^{\;-}(k,p_+) &=& \overline{V}_{--}^{\;+}(k,p_+)
  \label{36}\\
  &=& 
  \overline{V}_{++}^{\;0}(k,p_+) =
  \overline{V}_{--}^{\;0}(k,p_+)=0\,.
  \nn
\eea

The properties of the new vertices are quite similar to those for the
previous bremsstrahlung vertices. In particular, vertices with
the maximal change of helicity
\be
  \max |\Delta \lambda|= \max|\Lambda - \lambda_+ -\lambda_-|=2
  \label{38}
\ee
are absent. If the final $e^{\pm}$ becomes very hard ($x_{\pm} \to
1$, $x_{\mp} \to 0$) the initial photon ``transmits'' its helicity
to this hard lepton, i.e. the vertex $\overline{V}(k,p_+)$ with
$\Lambda = 2\lambda_{\pm}$ dominates at $x_{\pm} \to 1$. For
helicity non-conserved vertices there is a strong correlation
between the helicity of the photon and the positron:
\be
  \Lambda = 2 \lambda_+ \; \;\;\mbox{ if }\;\;\;
  \lambda_+=\lambda_-\,.
  \label{39}
\ee

\subsection{Substitution rules for the cross-channel}
\label{sec:2.3}

Let us consider the direct process of Fig.~\ref{fig:9}
 \be
a(p_1) + b(p_2) \to c(k)  + \ldots
 \ee
and the so-called conjugated or cross-process of Fig.~\ref{fig:10}
 \be
\bar{c}(\bar{p}_1)+ b(p_2)  \to \bar{a}(\bar{k})  + \ldots
 \ee
where $\bar{a}$ and $\bar{c}$ denote the corresponding
antiparticles. It is well-known that the amplitude
$\overline{M}_{fi}$ of the cross-channel can be obtained from the
amplitude ${M}_{fi}$ of the direct process by replacing:
 \be
p_1\to -\bar{k},\;\; k\to -\bar{p}_1\,,\;\;
 \lambda_a\to -\lambda_{\bar{a}}\,,\;\;
\lambda_c\to -\lambda_{\bar{c}}
 \label{crosshelicities}
 \ee
(and, may be, by an additional change of sign, cf. (\ref{22}) and
footnote 1.4). In this section the $\lambda_{a,c}$ generically
denote the helicities of both leptons and photons.
\begin{figure}[!htb]
  \centering
  \unitlength=2.00mm
  \begin{picture}(44.00,17.00)
    \put(23.00,12.00){\circle{4.00}}
    \put(23.00,12.00){\makebox(0,0)[cc]{$M_1$}}
    \put(23.00, 3.00){\circle{4.00}}
    \put(23.00, 3.00){\makebox(0,0)[cc]{$M_2$}}
    \put( 9.00, 3.00){\vector(1,0){6.50}}
    \put( 9.00,12.00){\vector(1,0){6.50}}
    \put(14.00, 3.00){\line(1,0){6.90}}
    \put(14.00,12.00){\line(1,0){6.90}}
    \put(25.00,12.50){\vector(1,0){6.00}}
    \put(25.00,11.50){\vector(1,0){6.00}}
    \put(24.50,13.50){\vector(3,1){6.50}}
    \put(23.00, 5.10){\line(0,1){0.90}}
    \put(23.00, 6.50){\line(0,1){2.00}}
    \put(23.00, 6.50){\vector(0,1){1.40}}
    \put(23.00, 9.00){\line(0,1){0.90}}
    \put(24.50,10.50){\vector(1,0){6.50}}
    \put(25.00, 3.50){\vector(1,0){6.00}}
    \put(25.00, 2.50){\vector(1,0){6.00}}
    \put(24.50, 4.50){\vector(1,0){6.50}}
    \put(24.50, 1.50){\vector(1,0){6.50}}
    \put(33.00,11.50){\makebox(0,0)[cc]{$p_i$}}
    \put(34.00,16.00){\makebox(0,0)[cc]{$c(k)$}}
    \put(21.00, 7.00){\makebox(0,0)[cc]{$q$}}
    \put( 5.00,12.00){\makebox(0,0)[cc]{$a(p_1)$}}
    \put( 5.00, 3.00){\makebox(0,0)[cc]{$b(p_2)$}}
    \put(39.00, 3.00){\makebox(0,0)[cc]{{\Huge \}}\ jet$_2$}}
    \put(39.00,13.00){\makebox(0,0)[cc]{{\Huge \}}\ jet$_1$}}
  \end{picture}
  \caption{Generic block diagram of the direct
 process  $a(p_1) +  b(p_2) \to c(k)+...\,+\, {\rm{jet}}_2$}
  \label{fig:9}
\end{figure}
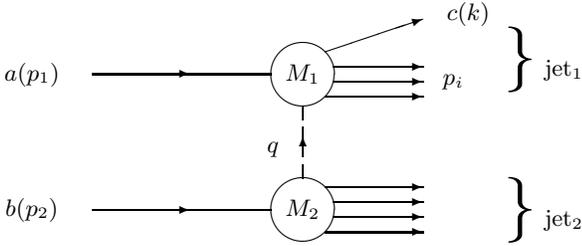
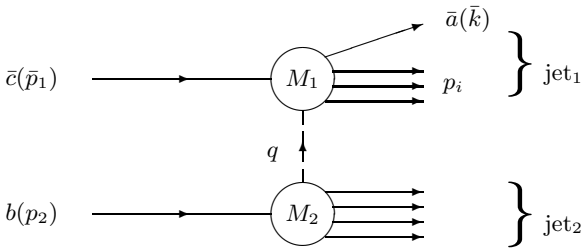
\begin{figure}[!htb]
  \centering
  \unitlength=2.00mm
  \begin{picture}(44.00,17.00)
    \put(23.00,12.00){\circle{4.00}}
    \put(23.00,12.00){\makebox(0,0)[cc]{$M_1$}}
    \put(23.00,3.00){\circle{4.00}}
    \put(23.00,3.00){\makebox(0,0)[cc]{$M_2$}}
    \put(9.00,3.00){\vector(1,0){6.50}}
    \put(9.00,12.00){\vector(1,0){6.50}}
    \put(14.00,3.00){\line(1,0){6.90}}
    \put(14.00,12.00){\line(1,0){6.90}}
    \put(25.00,12.50){\vector(1,0){6.00}}
    \put(25.00,11.50){\vector(1,0){6.00}}
    \put(24.50,13.50){\vector(3,1){6.50}}
    \put(24.50,10.50){\vector(1,0){6.50}}
    \put(23.00,5.10){\line(0,1){0.90}}
    \put(23.00,6.50){\line(0,1){2.00}}
    \put(23.00,6.50){\vector(0,1){1.40}}
    \put(23.00,9.00){\line(0,1){0.90}}
    \put(25.00,3.50){\vector(1,0){6.00}}
    \put(25.00,2.50){\vector(1,0){6.00}}
    \put(24.50,4.50){\vector(1,0){6.50}}
    \put(24.50,1.50){\vector(1,0){6.50}}
    \put(33.00,11.50){\makebox(0,0)[cc]{$p_i$}}
    \put(34.00,16.00){\makebox(0,0)[cc]{$\bar{a}(\bar{k})$}}
    \put(21.00,7.00){\makebox(0,0)[cc]{$q$}}
    \put(5.00,12.00){\makebox(0,0)[cc]{$\bar{c}(\bar{p}_1)$}}
    \put(5.00,3.00){\makebox(0,0)[cc]{$b(p_2)$}}
    \put(39.00,3.00){\makebox(0,0)[cc]{{\Huge \}}\ jet$_2$}}
    \put(39.00,13.00){\makebox(0,0)[cc]{{\Huge \}}\ jet$_1$}}
  \end{picture}
  \caption{ Generic block diagram of the cross-process  $\bar{c}(\bar{p}_1) +
  b(p_2) \to \bar{a}(\bar{k})+... \,+\, {\rm{jet}}_2$}
  \label{fig:10}
\end{figure}

If, for example, the considered initial particle is an electron,
$a=e^-(p_1)$, and  the final particle to be crossed a photon,
$c=\gamma(k)$, then $\bar{a}=e^+(\bar{k})$ and
$\bar{c}=\gamma(\bar{p}_1)$. The initial electron with 4-momentum
$p_1$ and helicity $\lambda_1$ is described by the bispinor
$u^{(\lambda_1)}_{{\vec p}_1}$ in the amplitude of the direct
process. In the conjugated process the particle $\bar{a}$ is the
final positron with 4-momentum $\bar{k}\equiv p_+$ and helicity
$\lambda_+$ described by the bispinor $v^{(\lambda_+)}_{{\vec
p}_+}$. These bispinors are connected by relation (\ref{23}).
Analogously, the final photon with 4-momentum $k$ and helicity
$\Lambda$ is described by the polarization vector
$e^{(\Lambda)*}(k)$ in $M_{fi}$. For the conjugated process in
$\overline{M}_{fi}$ the initial photon with 4-momentum
$\bar{p}_1=-k$ and helicity $\Lambda$ is described by
$e^{({\Lambda})}(-k)$. These polarization vectors are connected by
relation (\ref{22}).

Between the Sudakov variables for the direct and conjugated
processes certain relations exist which were obtained
in~\cite{KLS74}. In the direct process we introduce the Sudakov
variables for the initial particle $a$ with 4-momentum $p_1$, the
particle $c$ in the first jet with 4-momentum $k$ and for some
other particle in the first jet with $p_i$, using the
definition\footnote{Formulae given below are valid up to terms of
the relative order of $(m_i^2+{\vec p}_{i\perp}^2)/s$ or
$(\bar{m}_i^2+\bar{\vec p}^2_{i\perp})/ \bar{s}$ which we
systematically omit (here $s=2p_1p_2\,,\;\; \bar{s}=2\bar{p}_1
\bar{p}_2\,,\;\;\bar{p}_2=p_2$). }:
\be
  k=xP_1+yP_2+k_\perp\,,\;\; p_i=x_iP_1+y_iP_2+p_{i\perp}\,,
\ee
$$
 P_{1,2}=p_{1,2}-{p^2_{1,2}\over s}\, p_{2,1}\,,\;\;
 P_{1,2}^2=0\,,\;\; k_\perp p_{1,2}=p_{i\perp} p_{1,2}=0 \,.
$$
For the conjugated process we have
\be
 \bar k=\bar x \bar{P}_1+ \bar y \bar{P}_2+\bar{k}_\perp\,,\;\;
 {p}_i=\bar{x}_i \bar{P}_1+\bar{y}_i\bar{P}_2+\bar{p}_{i\perp}\,,
\ee
$$
 \bar{P}_{1,2}=\bar{p}_{1,2}-{\bar{p}^2_{1,2}\over \bar{s}}\,
 \bar{p}_{2,1}\,,\;\;
 \bar{P}_{1,2}^2=0\,,\;\; \bar{k}_\perp \bar{p}_{1,2}=\bar{p}_{i\perp}
  \bar{p}_{1,2}=0\,.
$$

In the jet-like kinematics all particles in the first jet have large
components along $P_1$ in the direct process (or along $\bar{P}_1$
in the conjugated process) and small components along $P_2$ (or
$\bar{P}_2$), therefore, $x,\,x_i,\,\bar x,\, \bar{x}_i \sim 1$,
while $y\sim (k^2+{\vec k}^2_{\perp})/s$, $y_i \sim (m_i^2+{\vec
p}^2_{i\perp})/s$ and $\bar y \sim (\bar{k}^2+\bar{\vec
k}^2_{\perp})/\bar{s}$, $\bar{y}_i \sim (\bar{m}_i^2+\bar{\vec
p}^2_{i\perp})/ \bar{s}$. Comparing
$$
  x={2kP_2\over s} = {kp_2\over p_1p_2}\,,\;\; x_i={p_ip_2\over
  p_1p_2}
$$
with
$$
  \bar{x}={2\bar{k} \bar{P}_2\over \bar{s}} = {p_1p_2\over
  kp_2}\,,\;\;\; \bar{x}_i=-{p_ip_2\over kp_2}\,,
$$
we immediately obtain the relations
\be
  x={1\over \bar{x}}\,, \;\;\; x_i = - {\bar{x}_i\over \bar{x}}\,,\;\;\;
  s=-\bar{s} \bar{x}\,.
  \label{43}
\ee

Presenting the equation $-p_1=\bar k=\bar x \bar{P}_1+ \bar y
\bar{P}_2+\bar{k}_\perp$ in the form
$$
  -p_1=-\bar x k+ \bar{b} p_2+\bar{k}_\perp\,,\;\;\bar{b}=\bar{y}-
  {\bar{x} k^2\over \bar{s}}\,,
$$
the 4-vector $k$ from the right-hand-side can be expressed as
follows
$$
  k= {1\over \bar{x}} (p_1 +\bar{b}p_2+\bar{k}_\perp)\,.
$$
Comparing this equation with
$$
  k=xP_1+yP_2+k_\perp=xp_1+ bp_2 +k_\perp\,, \;\; b=y-{xp_1^2\over
  s}\,,
$$
and taking into account (\ref{43}) and the conditions $k_\perp
p_2= \bar{k}_\perp p_2=0$, we obtain
\be
  k_\perp ={\bar{k}_\perp\over \bar{x}}\,.
  \label{44}
\ee

Performing a similar comparison for the 4-momenta of the other
particles in the first jet between
$$
  p_i=\bar{x}_i \bar{P}_1+\bar{y}_i\bar{P}_2+\bar{p}_{i\perp}=
  -\bar{x}_i x p_1+ \bar{b}_i p_2 - \bar{x}_i k_\perp +\bar{p}_{i\perp}
$$
and
$$
  p_i=x_i P_1+y_i P_2+p_{i\perp}=x_i p_1+ b_i p_2 +p_{i\perp}\,,
$$
we find another useful relation
\be
  p_{i\perp}= \bar{p}_{i\perp}- {\bar{x}_i\over \bar{x}}\,
  \bar{k}_\perp\,.
  \label{45}
\ee

As expected, the transformations (\ref{43}) respect the
conservation of energy fraction  and (\ref{44}, \ref{45}) the
transverse momentum conservation of the direct and conjugated process
within the first jet:
\bea
  && 1=x+\sum_i x_i=\bar{x}+\sum_i \bar{x}_i\,, 
  \nn\\
  && q_\perp=k_\perp+\sum_i k_{i\perp}=\bar{k}_\perp+\sum_i \bar{k}_{i\perp}
  \,.
  \nn
\eea
Since the impact factor $J_1$ of that jet depends only on energy
fractions, transverse momenta and helicities of all final
particles in  the jet and of the incoming particle (energy
fraction equal to one, no transverse momentum), the
transformations (\ref{43}-\ref{45}) together with the helicity
changes (\ref{crosshelicities}) for the particles to be crossed
completely describe the transformation from an impact factor in
the direct process to that of the cross-process. Finally, a
possible sign change in the impact factor as a result of
transformation of the form (\ref{22}) has to be taken into
account.

\section{Pair production in $\gamma e$ and $\gamma \gamma$ collisions}
\label{sec:3}

The impact factor for the single lepton pair production of 
Fig.~1.3 can be obtained from (1.61) by the replacements
\bea
  &&  p_1 \to -p_+\,,\; p_3 \to p_-\,,\;   k\to -k\,,
  \nn\\
  && e^{(\Lambda)*}(k) \to e^{(\Lambda)}(k)\,,\;\; 
  u_1 \to v_+\,,\;\; u_3 \to u_-
  \label{46}
\eea
and is of the form
\be
  J_1 ( \gamma_\Lambda +\gamma^* \to e^+_{\lambda_+}+e^-_{\lambda_-})=
  4\pi\alpha\, \left(\frac{N_+}{2kp_+} + \frac{N_-}{2kp_-}\right)
  \label{47}
\ee
\bea
  N_+&=& \bar u_- \, \hat{e}_q \, (-\hat{p}_+ +\hat k +m) \, \hat{e}\, v_+\,,
  \nn\\
  N_-&=& \bar u_- \, \hat{e} \, (\hat{p}_- -\hat k +m) \, \hat{e}_q \, v_+\,,
  \nn
\eea
$e \equiv e^{(\Lambda)}(k)$ is the polarization 4-vector of
the initial photon and $e_q=\sqrt{2} P_2/s$. Following along the
lepton line and  using the rules from Sect.~\ref{sec:2.1}, the impact factor
can be written in the form
\bea
 J_1=
  4\pi \alpha&&
  \Bigg[
  \frac{\overline{V}_{\lambda_+\, \lambda}^{\;\Lambda}(k,p_+)
   V_{\lambda\, \lambda_-}(k-p_+)}{2kp_+}
  \nn\\
  && -
  \frac{\overline{V}_{\lambda_-\, \lambda}^{\;\Lambda}(k,p_-)
  V_{\lambda\, \lambda_+}(k-p_-)}{2kp_-}
  \Bigg] \,,
  \label{getoe+e_e}
\eea
where the vertices $V(p)$ and $\overline{V}(k,p_{\pm})$ are given
by (\ref{15}) and (\ref{25}), (\ref{29}), respectively.

In the  following we could  repeat here all steps of the
calculation as described in Sect.~4 of~\cite{CSS1}. However,
instead of performing the calculations, the final expression can
be directly obtained from (1.75), (1.76) using the substitution
rules:
\bea
  &&J_1(e\gamma^* \to e\gamma) \to -{J_1(\gamma \gamma^* \to
  e^+e^-)\over x_+}
  \nn\\
  &&x\to {1\over x_+}\,,\; {k_\perp\over x} \to p_{+\perp}\,,\;\;
  q_\perp-{k_\perp\over x} \to p_{-\perp}\,,  
  \label{49}\\
  && \lambda_1  \to  -\lambda_+\,,\;
  \lambda_3 \to \lambda_-,\;\Lambda\to -\Lambda\,,\;
  \varphi_1 \to \varphi_+\,,\;\varphi_3 \to \varphi_-
  \nn
\eea
where $x_{\pm} = E_{\pm}/\omega$. This gives
\bea
  &&J_1 ( \gamma_\Lambda +\gamma^* \to
  e^+_{\lambda_+}+e^-_{\lambda_-})= 
  \nn\\
   &&=8\pi\alpha\;{\rm i} \sqrt{x_+ x_-}\, {\rm e}^{{\rm i}
  (\lambda_+ \varphi_+ + \lambda_- \varphi_-)} \times
  \label{50} \\
  &&
  \left[ \left(x_+ - \delta_{\Lambda,- 2\lambda_+}\right)
  \sqrt{2} {\vec T} {\vec e}^{(\Lambda)}_\perp 
  \delta_{\lambda_+,- \lambda_-}-
  m S \delta_{\lambda_+\lambda_-}
  \delta_{\Lambda, 2 \lambda_+} \right] \,.
  \nn
\eea
The transverse 4-vector $T$ (in the used reference frame
$T=(0,\vec T,0), \, \, T^2=-{\vec T}^2$) and the scalar $S$
are defined as
\bea
  && 
  T={p_{+\perp} \over a} + {p_{-\perp}\over b}\,, \;\;
  S={1\over a} - {1\over b}\,,
  \nn\\
  &&
  a=m^2+{\vec p}^2_{+\perp}\,,\;\; 
  b=m^2+{\vec p}^2_{-\perp}\,,
 \label{51}
\eea
with the useful relation
\be
  {\vec T}^2 + m^2 S^2 = {{\vec q}_\perp^2 \over ab}\,.
  \label{52}
\ee

Since $T \propto q_\perp$ and $S\propto q_\perp$, we conclude that
$J_1 \propto q_\perp$. The impact factor (\ref{getoe+e_e}),
(\ref{50}) changes its sign under the exchange $+ \leftrightarrow
-$ which is directly related to the obvious symmetry of diagrams
of Fig.~1.3 under lepton exchange $e^+ \leftrightarrow e^-$.

It is interesting to compare the structure of the amplitudes
obtained for single bremsstrahlung (Fig.~1.2) and for single pair
production (Fig.~1.3). In the former case using (1.63), we can
express the amplitude of the inelastic process $ee \to ee \gamma$
via the amplitudes of {\it one} elastic process $ee \to ee$ but
with different momenta for leptons of the upper block:
\bea
  &&M_{ee\to ee\gamma}= \sqrt{4\pi \alpha}
  \nn\\
  &&\Bigg\{
  \frac{V_{\lambda_1\,\lambda}^{\;\Lambda}(p_1,k)}{ 2kp_1}
  \,\left[M_{ee\to ee}(p_1-k,p_3)\right]_{\lambda\, \lambda_3}
   \nn\\
  &&
  -\left[M_{ee\to ee}(p_1, p_3+k)\right]_{\lambda_1\,
  \lambda}
  \frac{V_{\lambda\,\lambda_3}^{\;\Lambda}(p_3+k,k)}{2kp_3}\,\Bigg\}\,.
  \label{53}
\eea
This expression can be considered as a generalization of the
well-known classical current approximation,
\be
M_{ee\to ee\gamma}= \sqrt{4\pi \alpha}\, \left[ {p_1e^*\over
kp_1}- {p_3e^*\over kp_3}\right] \,M_{ee\to ee}(p_1,p_3)\,,
 \label{54}
\ee
which is valid for soft photon emission, $\omega\ll E_{1,3}$.
On the other hand, Eq. (\ref{53}) is valid for the emission of a photon
with arbitrary energy but for small angles of the final photon
and lepton.

For single pair production an equation analogous to (\ref{53}) can
be directly derived from (\ref{getoe+e_e}):
\bea
 && M_{\gamma e\to l^+l^- e}= \sqrt{4\pi\alpha}
\nn\\
&&
  \left\{{\overline{V}_{\lambda_+ \, \lambda}^{\;\Lambda}(k,p_+)\over 2kp_+}\,
  \left[M_{l^-e\to l^-e}(k-p_+,p_-)\right]_{\lambda\, \lambda_-} \right.
  \nn\\
   &&\left. +
  {\overline{V}_{\lambda_-\, \lambda}^{\;\Lambda}(k, p_-)\over 2kp_-}\,
  \left[M_{l^+e\to l^+e}(k-p_-,p_+)\right]_{\lambda\, \lambda_+}
  \,\right\}\,.
 \label{55}
\eea
In contrast to the bremsstrahlung case, the amplitude for pair
production is expressed via amplitudes for {\it two different}
elastic processes: $l^+ e \to l^+ e$ and $l^- e \to l^- e$.

Having the impact factors for single bremsstrahlung (1.68) or
(1.75) and for single pair production (\ref{50}), we are able to
obtain the amplitudes for the processes of Figs.~1.2--1.6. More details
about the double bremsstrahlung in opposite directions of Fig.~1.4 and
the double pair production of Fig.~1.5 can be found in 
Refs.~\cite{kuraevzp86} and \cite{kuraevnp85}, respectively.  In the
same manner, using the impact factor for double bremsstrahlung
along one direction from Sect.~5 of \cite{CSS1}, we get the
impact factor for the process of Fig.~1.10, the corresponding
calculations can be found in Ref.~\cite{KSSSh00}.

\section{Pair production in $e^+e^-$ collisions}
\label{sec:4}

We start with bremsstrahlung $\mu^+ \mu^-$ pair production of
Fig.~1.8. The corresponding impact factor is given by the expression (see
Sect.~\ref{sec:2.1})
\bea
  &&J_1 (e^-_{\lambda_1}+\gamma^* \to e^-_{\lambda_3}+
  \mu^+_{\lambda_+} +\mu^-_{\lambda_-})
  \nn\\
  && = - {\sqrt{4\pi\alpha}\over k^2}\, 
  \sum_{\Lambda=0,\pm 1}\,
   A^{(\Lambda)}\overline{V}^{\,\Lambda}(k,p_+)\,.
  \label{56}
\eea
The vertex $\overline{V}(k,p_+)$ has been obtained already in
(\ref{25}) (the mass has to be identified with the muon mass
$m_\mu$), the virtuality of the photon $k^2=(p_+ + p_-)^2$ is
given via the energy fractions $x_\pm=E_{\pm}/E_1$ (with $x_+
+x_-=x=\omega/E_1$) and transverse momenta $\vec p_\pm$ (with $\vec
p_{+\perp}+\vec p_{-\perp}=\vec k_\perp$) of the muons:
\be
  k^2= \frac{1}{x_+ x_-} \left[ x^2 m_\mu^2 +
 \left(x_- {\vec p}_{+\perp} - x_+ {\vec p}_{-\perp}\right)^2 \right]\,.
 \label{virtk2}
\ee

Therefore, it remains to calculate the quantities $A^{(\pm)}$ and
$A^{(0)}$ only.
Using the rules from Sect.~\ref{sec:2.1}, we obtain the expression similar
to (1.63):
\bea
  A^{(\Lambda)}=
  4\pi \alpha\,
  \left[
  \frac{V(p_1,k) V(p_1-k)}{2kp_1-k^2}-
  \frac{V(p_1)V(p_3+k,k)}{2kp_3+k^2}
  \right] \,,
\nn
\eea
the vertices $V(p)$ and $V(p,k)$ are given by (\ref{15}) and
(\ref{16}), respectively.
Since
\bea
  && 2kp_1-k^2 = x a\,, \quad a=\,m^2 + {{\vec k}^2_\perp \over x^2}+
  {1-x\over x^2}k^2\,,
  \nn\\
  && 2kp_3+k^2 = {x\over 1-x}\,b\,, \;\; 
  \label{58}\\
  && b=m^2 +
  \left({\vec q}_\perp
  - {{\vec k}_\perp \over x}\right)^2+{1-x\over x^2}k^2\,,
  \nn
\eea
where $m$ is the electron mass, we can repeat the derivation of
(1.75) with the result
\bea
  &&
  A^{(\Lambda)}=8\pi\alpha {\sqrt{1-x}\over x}\, {\rm e}^{{\rm
  i} (\lambda_3 \varphi_3 - \lambda_1 \varphi_1)} 
  \label{61} \\
  &&\times \left[ \left(1- x \delta_{\Lambda,- 2\lambda_1}\right) 
  \sqrt{2} {\vec T} {\vec e}^{(\Lambda)\,*}_\perp \,
  \delta_{\lambda_1 \lambda_3}+
  m \,x S \delta_{\lambda_1, -\lambda_3}
  \delta_{\Lambda, 2 \lambda_1} \right]
 \nn
 \eea
for $\Lambda=\pm 1$ and
\be
  A^{(0)}=8\pi\alpha {\sqrt{1-x}\over x}\, {\rm e}^{{\rm i}
  (\lambda_3 \varphi_3 - \lambda_1 \varphi_1)} \; {\sqrt{k^2}\over
  x}\, S \, \delta_{\lambda_1 \lambda_3}
  \label{61a}
\ee
for $\Lambda=0$.
Here we have used the scalar $S$ and the transverse 4-vector $T$:
\be
  S= {1\over a} - {1\over b}\,,\;\;
  T={(k_\perp/x) \over a} + {q_\perp -(k_\perp/x)\over b}\,.
  \label{59}
 \ee
They obey the relation
\be
  {\vec T}^2 + \left( m^2 +{1-x\over x^2}k^2 \right)\,S^2 = {{\vec
  q}_\perp^2 \over ab}
  \label{63}
\ee
from which it follows that $J_1 \propto q_\perp$.

The final results (\ref{56}), (\ref{61}), (\ref{61a}) and
(\ref{25}) coincide with those obtained in~\cite{KSSS98} by means
of considerably more complicated calculations. The impact factor
for the cross-channel of Fig.~1.7 can be obtained using the
substitution rules, the corresponding expression is given
in~\cite{KSSS98}.

\section{Bremsstrahlung $\mu^+ \mu^-$ pair production with
additional photon} 
\label{sec:5}

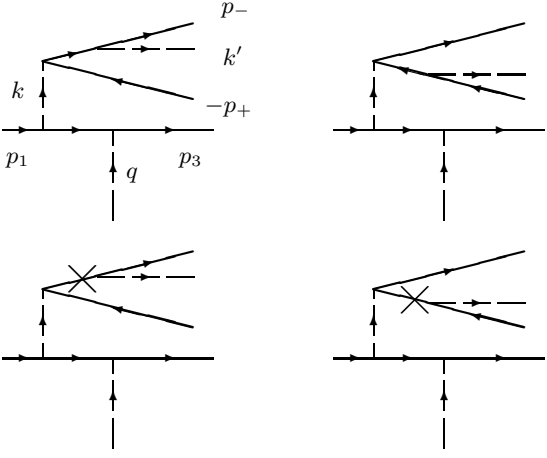
\begin{figure}[!htb]
  \centering
  \unitlength=2.00mm
  \begin{picture}(40.00,15.50)
   \put( 1.00, 6.00){\line(1,0){ 7.0}}
   \put( 1.00, 6.00){\vector(1,0){1.80}}
   \put( 4.50, 6.00){\vector(1,0){1.80}}
   \put( 2.00, 4.00){\makebox(0,0)[cc]{$p_1$}}
   \put( 8.00, 6.00){\line(1,0){ 7.0}}
   \put( 8.00, 6.00){\vector(1,0){4.50}}
   \put(13.50, 4.00){\makebox(0,0)[cc]{$p_3$}}
   \put( 2.00, 8.70){\makebox(0,0)[cc]{$k$}}
   \put( 9.60, 3.00){\makebox(0,0)[cc]{$q$}}
   \put( 8.34, 0.00){\line(0,1){2.00}}
   \put( 8.34, 2.50){\line(0,1){2.00}}
   \put( 8.34, 5.00){\line(0,1){1.00}}
   \put( 8.34, 2.50){\vector(0,1){1.40}}
   \put( 3.66,10.00){\line(0,1){0.60}}
   \put( 3.66, 7.50){\line(0,1){2.00}}
   \put( 3.66, 6.00){\line(0,1){1.00}}
   \put( 3.66, 7.50){\vector(0,1){1.40}}
   \put(13.66, 8.05){\vector(-4,1){5.4}}
   \put( 3.66,10.60){\vector(4,1){7.4}}
   \put( 3.66,10.60){\vector(4,1){2.4}}
  {\thicklines
   \put( 3.66,10.60){\line(4, 1){10.0}}
   \put( 3.66,10.60){\line(4,-1){10.0}}
  }
   \put(16.50,14.00){\makebox(0,0)[cc]{$p_-$}}
   \put(16.00, 7.50){\makebox(0,0)[cc]{$-p_+$}}
   \put( 7.30,11.40){\line(1,0){1.80}}
   \put( 9.60,11.40){\line(1,0){1.80}}
   \put( 9.60,11.40){\vector(1,0){1.30}}
   \put(11.90,11.40){\line(1,0){1.80}}
   \put(16.30,11.00){\makebox(0,0)[cc]{$k'$}}
   \put(23.00, 6.00){\line(1,0){ 7.0}}
   \put(23.00, 6.00){\vector(1,0){1.80}}
   \put(26.50, 6.00){\vector(1,0){1.80}}
   \put(30.00, 6.00){\line(1,0){ 7.0}}
   \put(30.00, 6.00){\vector(1,0){4.50}}
   \put(30.34, 0.00){\line(0,1){2.00}}
   \put(30.34, 2.50){\line(0,1){2.00}}
   \put(30.34, 5.00){\line(0,1){1.00}}
   \put(30.34, 2.50){\vector(0,1){1.40}}
   \put(25.66,10.00){\line(0,1){0.60}}
   \put(25.66, 7.50){\line(0,1){2.00}}
   \put(25.66, 6.00){\line(0,1){1.00}}
   \put(25.66, 7.50){\vector(0,1){1.40}}
   \put(35.66, 8.05){\vector(-4,1){8.4}}
   \put(35.66, 8.05){\vector(-4,1){3.6}}
   \put(25.66,10.60){\vector(4,1){5.4}}
  {\thicklines
   \put(25.66,10.60){\line(4, 1){10.0}}
   \put(25.66,10.60){\line(4,-1){10.0}}
  }
   \put(29.30, 9.70){\line(1,0){1.80}}
   \put(31.60, 9.70){\line(1,0){1.80}}
   \put(31.60, 9.70){\vector(1,0){1.30}}
   \put(33.90, 9.70){\line(1,0){1.80}}
  \end{picture}
  \begin{picture}(40.00,15.00)
   \put( 1.00, 6.00){\line(1,0){ 7.0}}
   \put( 1.00, 6.00){\vector(1,0){1.80}}
   \put( 4.50, 6.00){\vector(1,0){1.80}}
   \put( 8.00, 6.00){\line(1,0){ 7.0}}
   \put( 8.00, 6.00){\vector(1,0){4.50}}
   \put( 8.34, 0.00){\line(0,1){2.00}}
   \put( 8.34, 2.50){\line(0,1){2.00}}
   \put( 8.34, 5.00){\line(0,1){1.00}}
   \put( 8.34, 2.50){\vector(0,1){1.40}}
   \put( 3.66,10.00){\line(0,1){0.60}}
   \put( 3.66, 7.50){\line(0,1){2.00}}
   \put( 3.66, 6.00){\line(0,1){1.00}}
   \put( 3.66, 7.50){\vector(0,1){1.40}}
   \put(13.66, 8.05){\vector(-4,1){5.4}}
   \put( 3.66,10.60){\vector(4,1){7.4}}
   \put( 5.40,10.40){\line(1,1){1.80}}
   \put( 5.40,12.20){\line(1,-1){1.80}}
  {\thicklines
   \put( 3.66,10.60){\line(4, 1){10.0}}
   \put( 3.66,10.60){\line(4,-1){10.0}}
  }
   \put( 7.30,11.40){\line(1,0){1.80}}
   \put( 9.60,11.40){\line(1,0){1.80}}
   \put( 9.60,11.40){\vector(1,0){1.30}}
   \put(11.90,11.40){\line(1,0){1.80}}
   \put(23.00, 6.00){\line(1,0){ 7.0}}
   \put(23.00, 6.00){\vector(1,0){1.80}}
   \put(26.50, 6.00){\vector(1,0){1.80}}
   \put(30.00, 6.00){\line(1,0){ 7.0}}
   \put(30.00, 6.00){\vector(1,0){4.50}}
   \put(30.34, 0.00){\line(0,1){2.00}}
   \put(30.34, 2.50){\line(0,1){2.00}}
   \put(30.34, 5.00){\line(0,1){1.00}}
   \put(30.34, 2.50){\vector(0,1){1.40}}
   \put(25.66,10.00){\line(0,1){0.60}}
   \put(25.66, 7.50){\line(0,1){2.00}}
   \put(25.66, 6.00){\line(0,1){1.00}}
   \put(25.66, 7.50){\vector(0,1){1.40}}
   \put(35.66, 8.05){\vector(-4,1){3.6}}
   \put(25.66,10.60){\vector(4,1){5.4}}
   \put(27.50, 9.00){\line(1,1){1.80}}
   \put(27.50,10.80){\line(1,-1){1.80}}
  {\thicklines
   \put(25.66,10.60){\line(4, 1){10.0}}
   \put(25.66,10.60){\line(4,-1){10.0}}
  }
   \put(29.30, 9.70){\line(1,0){1.80}}
   \put(31.60, 9.70){\line(1,0){1.80}}
   \put(31.60, 9.70){\vector(1,0){1.30}}
   \put(33.90, 9.70){\line(1,0){1.80}}
  \end{picture}
  \caption{Diagrams
  (including crossed intermediate muon lines)
   for the impact factor related to
   bremsstrahlung $\mu^+ \mu^-$ pair production with an
   additional photon emitted by muons}
  \label{fig:11}
\end{figure}

\begin{figure}[!htb]
  \begin{center}
  \unitlength=2.00mm
  \begin{picture}(41.00,16.00)
    \put( 1.00, 6.00){\line(1,0){16.0}}
    \put( 1.00, 6.00){\vector(1,0){1.50}}
    \put( 5.80, 6.00){\vector(1,0){2.90}}
    \put( 7.00, 6.00){\vector(1,0){9.0}}
    \put(12.40, 6.00){\vector(1,0){1.0}}
    \put( 2.00, 4.50){\makebox(0,0)[cc]{$p_1$}}
    \put(17.00, 4.50){\makebox(0,0)[cc]{$p_3$}}
    \put(14.00, 0.00){\line(0,1){2.00}}
    \put(14.00, 2.50){\line(0,1){2.00}}
    \put(14.00, 5.00){\line(0,1){1.00}}
    \put(14.00, 2.50){\vector(0,1){1.40}}
    \put(12.50, 3.00){\makebox(0,0)[cc]{$q$}}
    \put(14.00, 8.50){\line(1,1){1.8}}
    \put(14.00, 8.50){\vector(1,1){1.2}}
    \put(11.50, 6.00){\line(1,1){1.8}}
    \put(18.00, 9.00){\makebox(0,0)[cc]{$k_2$}}
    \put( 9.00,11.00){\line(1,1){1.8}}
    \put( 6.50, 8.50){\line(1,1){1.8}}
    \put( 4.00, 6.00){\line(1,1){1.8}}
    \put( 6.50, 8.50){\vector(1,1){1.2}}
    \put( 5.00, 9.50){\makebox(0,0)[cc]{$k_1$}}
   {\thicklines
    \put(10.80,12.80){\line(4,1){5.0}}
    \put(10.80,12.80){\line(4,-1){5.0}}}
    \put(15.80,11.60){\vector(-4,1){2.9}}
    \put(10.80,12.80){\vector(4,1){2.9}}
    \put(18.80,11.50){\makebox(0,0)[cc]{$p_-$}}
    \put(18.30,14.00){\makebox(0,0)[cc]{$-p_+$}}
    \put(23.00, 6.00){\line(1,0){16.0}}
    \put(23.00, 6.00){\vector(1,0){2.50}}
    \put(23.00, 6.00){\vector(1,0){6.50}}
    \put(23.00, 6.00){\vector(1,0){10.50}}
    \put(23.00, 6.00){\vector(1,0){14.50}}
    \put(31.00, 0.00){\line(0,1){2.00}}
    \put(31.00, 2.50){\line(0,1){2.00}}
    \put(31.00, 5.00){\line(0,1){1.00}}
    \put(31.00, 2.50){\vector(0,1){1.40}}
    \put(32.00,11.00){\line(1,1){1.8}}
    \put(29.50, 8.50){\line(1,1){1.8}}
    \put(27.00, 6.00){\line(1,1){1.8}}
    \put(29.50, 8.50){\vector(1,1){1.2}}
    \put(37.50, 8.50){\line(1,1){1.8}}
    \put(35.00, 6.00){\line(1,1){1.8}}
    \put(37.50, 8.50){\vector(1,1){1.2}}
   {\thicklines
    \put(33.80,12.80){\line(4,1){5.0}}
    \put(33.80,12.80){\line(4,-1){5.0}}}
    \put(38.80,11.60){\vector(-4,1){2.9}}
    \put(33.80,12.80){\vector(4,1){2.9}}
  \end{picture}
  \begin{picture}(41.00,16.00)
    \put(12.00, 6.00){\line(1,0){16.0}}
    \put(12.00, 6.00){\vector(1,0){1.50}}
    \put(15.66, 6.00){\vector(1,0){0.90}}
    \put(18.00, 6.00){\vector(1,0){3.0}}
    \put(23.40, 6.00){\vector(1,0){4.0}}
    \put(15.00, 0.00){\line(0,1){2.00}}
    \put(15.00, 2.50){\line(0,1){2.00}}
    \put(15.00, 5.00){\line(0,1){1.00}}
    \put(15.00, 2.50){\vector(0,1){1.40}}
    \put(22.00,11.00){\line(1,1){1.8}}
    \put(19.50, 8.50){\line(1,1){1.8}}
    \put(17.00, 6.00){\line(1,1){1.8}}
    \put(19.50, 8.50){\vector(1,1){1.8}}
    \put(26.50, 8.50){\line(1,1){1.8}}
    \put(24.00, 6.00){\line(1,1){1.8}}
    \put(26.50, 8.50){\vector(1,1){1.2}}
   {\thicklines
    \put(23.80,12.80){\line(4,1){5.0}}
    \put(23.80,12.80){\line(4,-1){5.0}}}
    \put(28.80,11.60){\vector(-4,1){2.9}}
    \put(23.80,12.80){\vector(4,1){2.9}}
  \end{picture}
  \begin{picture}(41.00,16.00)
    \put( 1.00, 6.00){\line(1,0){16.0}}
    \put( 1.00, 6.00){\vector(1,0){1.50}}
    \put( 7.00, 6.00){\vector(1,0){9.0}}
    \put(12.40, 6.00){\vector(1,0){1.0}}
    \put(14.00, 0.00){\line(0,1){2.00}}
    \put(14.00, 2.50){\line(0,1){2.00}}
    \put(14.00, 5.00){\line(0,1){1.00}}
    \put(14.00, 2.50){\vector(0,1){1.40}}
    \put(14.00, 8.50){\line(1,1){1.8}}
    \put(14.00, 8.50){\vector(1,1){1.2}}
    \put(11.50, 6.00){\line(1,1){1.8}}
    \put( 9.00,11.00){\line(1,1){1.8}}
    \put( 6.50, 8.50){\line(1,1){1.8}}
    \put( 4.00, 6.00){\line(1,1){1.8}}
    \put( 6.50, 8.50){\vector(1,1){1.2}}
   {\thicklines
    \put(10.80,12.80){\line(4,1){5.0}}
    \put(10.80,12.80){\line(4,-1){5.0}}}
    \put(15.80,11.60){\vector(-4,1){2.9}}
    \put(10.80,12.80){\vector(4,1){2.9}}
    \put( 7.00, 5.10){\line(1,1){1.80}}
    \put( 7.00, 6.90){\line(1,-1){1.80}}
    \put(23.00, 6.00){\line(1,0){16.0}}
    \put(23.00, 6.00){\vector(1,0){1.50}}
    \put(26.66, 6.00){\vector(1,0){0.90}}
    \put(34.40, 6.00){\vector(1,0){4.0}}
    \put(26.00, 0.00){\line(0,1){2.00}}
    \put(26.00, 2.50){\line(0,1){2.00}}
    \put(26.00, 5.00){\line(0,1){1.00}}
    \put(26.00, 2.50){\vector(0,1){1.40}}
    \put(33.00,11.00){\line(1,1){1.8}}
    \put(30.50, 8.50){\line(1,1){1.8}}
    \put(28.00, 6.00){\line(1,1){1.8}}
    \put(30.50, 8.50){\vector(1,1){1.8}}
    \put(37.50, 8.50){\line(1,1){1.8}}
    \put(35.00, 6.00){\line(1,1){1.8}}
    \put(37.50, 8.50){\vector(1,1){1.2}}
   {\thicklines
    \put(34.80,12.80){\line(4,1){5.0}}
    \put(34.80,12.80){\line(4,-1){5.0}}}
    \put(39.80,11.60){\vector(-4,1){2.9}}
    \put(34.80,12.80){\vector(4,1){2.9}}
    \put(30.50, 5.10){\line(1,1){1.80}}
    \put(30.50, 6.90){\line(1,-1){1.80}}
  \end{picture}
  \end{center}
  \caption{Diagrams for the impact factor related to
  bremsstrahlung $\mu^+ \mu^-$ pair production with
  an additional photon emitted by the electron
  }
  \label{fig:12}
\end{figure}
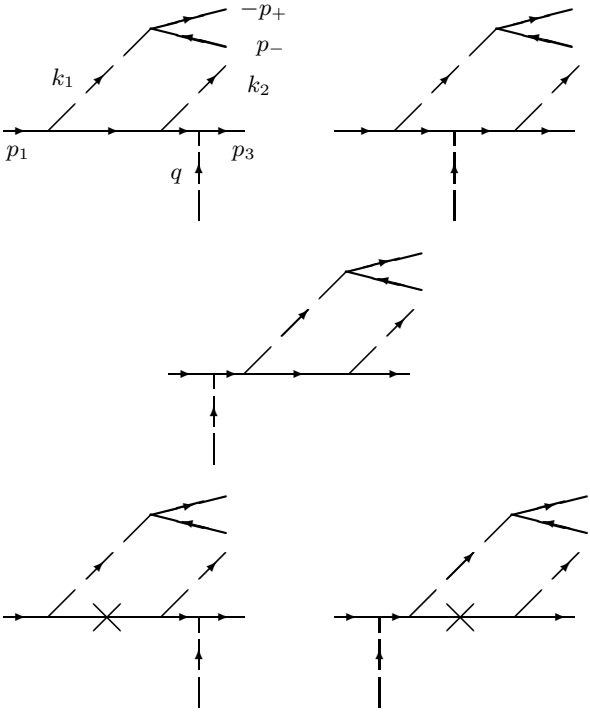

In this section we demonstrate how the newly defined rules in the
jet-like kinematics can be efficiently applied to the more complicated
reaction
$$
  ee \to ee \mu^+ \mu^- \gamma
$$ 
where both the produced muon pair and the photon belong to the
first jet. That process (which is shown schematically in
Fig.~\ref{fig:1}) is described by two sets of Feynman diagrams.
The first set is the bremsstrahlung $\mu^+ \mu^-$ pair production
of Fig.~1.8 with an additional photon line attached to the initial
electron line and to every muon and electron line in the first
jet. The second set is the two-photon $\mu^+ \mu^-$ pair
production of Fig.~1.7 adding again a photon line to both the
initial electron  and all final leptons line in the first jet. In
this section we consider the first set of diagrams having in mind
that the amplitudes for the second set can be obtained by a simple
transition to the cross-channel. The impact factor $J_1$ for the
set under discussion is described by the diagrams of 
Figs.~\ref{fig:11} and \ref{fig:12}.

\subsection{Photon emission by muons (Fig.~\ref{fig:11})}
\label{sec:5.1}

The contribution of Fig.~\ref{fig:11} can be written in the form
\bea
  &&J_1^{(\ref{fig:11})} (e^-_{\lambda_1}+\gamma^* \to e^-_{\lambda_3}+
  \mu^+_{\lambda_+} +\mu^-_{\lambda_-}+\gamma_{\Lambda'})
  \nn\\
  &&=
  - \frac{1}{k^2}\, A_\mu B^\mu=\frac{1}{k^2}\,\sum_{\Lambda=0,\pm 1}
  A^{(\Lambda)} B^{(\Lambda)}\,,
  \label{64}
\eea
where $A^{(\Lambda)}$ is given by (\ref{61}), (\ref{61a}) and
$B^{(\Lambda)}$ corresponds to the process $\gamma(k) \to
\mu^+(p_+) + \mu^-(p_-)+ \gamma(k')$:
\be
  B^{(\Lambda)} =  4\pi\alpha\left(\frac{N_-}{2p_-k'} -
  \frac{N_+}{2p_+k'}\right)\,
  \label{65}
\ee
with 
\bea 
  N_-&=& \bar{u}_- \hat{e}^{(\Lambda')*}(k')\,
  (\hat{p}_-+\hat{k}' +m_\mu)\,\hat{e}^{(\Lambda)}(k)\, v_+\,,
  \nn\\
  N_+&=& \bar{u}_- \hat{e}^{(\Lambda)}(k)\,
  (\hat{p}_++\hat{k}' -m_\mu)\,\hat{e}^{(\Lambda')*}(k')\, v_+\,.
  \nn
\eea
Here $m_\mu$ is the muon mass and
\bea
  && 2p_{\pm}k'= {x'\over x_{\pm}}\left( m_\mu^2 +{\vec
  r}^2_{\pm}\right)\,,\;\;
  {\vec r}_{\pm}= {\vec p}_{\pm
  \perp}-{x_{\pm}\over x'} {\vec k}'_\perp \,,
  \nn\\
  && x_{\pm}= {E_{\pm}\over E_1}\,,\;\; x'= {\omega'\over E_1}\,,
  \label{66c}\\
  && k^2= \frac{1}{x_+ x_-}\,\left[ \left(x_+ +x_-\right)^2 m_\mu^2
   + \left( x_- {\vec p}_{+\perp} - x_+ {\vec p}_{-\perp} \right)^2 \right]
  \nn\\
  &&+
  \frac{x'}{x_+}\,\left( m_\mu^2 +{\vec r}^2_{+}\right)+
  \frac{x'}{x_-}\,\left( m_\mu^2 +{\vec r}^2_{-}\right)\,.
  \label{66}
\eea
Using the vertices from Sect.~\ref{sec:2.1}, we immediately obtain
\bea
  {B^{(\Lambda)}\over 4\pi\alpha}&=& {1\over 2p_-k'}
  \overline{V}^{\,\Lambda}_{\lambda_+\,\lambda}(k,p_+)\,
  {V}^{\Lambda'}_{\lambda\,\lambda_-}(k-p_+,k')
  \nn\\
  &-& {1\over 2p_+k'}
  \overline{V}^{\,\Lambda}_{\lambda_-\,\lambda}(k,p_-)\,
  {V}^{\Lambda'}_{\lambda\,\lambda_+}(k-p_-,k')
  \nn\\
  &+&\overline{V}^{\,\Lambda\,\Lambda'}_{\lambda_+\,\lambda_-}(k,p_+,k')-
  \overline{V}^{\,\Lambda\,\Lambda'}_{\lambda_-\,\lambda_+}(k,p_-,k')
  \,.
  \label{67}
\eea
The last two items are due to diagrams of Fig.~\ref{fig:11} with
the crossed intermediate muon line.

For a soft final photon  ($x'\ll 1$, ${\vec k}'_\perp \to 0$ and
${\vec k}'_\perp/x'$ remains finite) we have
\be
  B^{(\Lambda)}= 4\pi\alpha\,
  \overline{V}^{\,\Lambda}_{\lambda_+\,\lambda_-}(k,p_+)\,
  \left({e^{'*}p_-\over
  p_-k'}-{e^{'*}p_+\over p_+k'} \right)\,.
\ee
This expression corresponds to the approximation of classical
currents. In that limit the virtuality $k^2$ of (\ref{66})
simplifies to (\ref{virtk2}).

\subsection{Photon emission by electrons (Fig.~\ref{fig:12})}
\label{sec:5.2}

It is convenient to use the same notation as for the process of
double bremsstrahlung (see Fig.~1.13), but to take into account that the
photon $k_1$ is now virtual with $k_1^2 > 0$ and its helicity
is $\Lambda_1=0,\pm 1$. In particular, we introduce the energy
fractions
\bea
  x_{1} =\frac{\omega_{1}}{E_1}&=&\frac{E_++E_-}{E_1}, \;\;
  x_{2} =\frac{\omega_{2}}{E_1}\,,
  \;\; X_3 = \frac{E_3}{E_1}\,,
  \nn\\
  && x_1+x_2+X_3=1\,.
  \nn
\eea
The denominators of the propagators in Fig.~\ref{fig:12} are
expressed via the energy fractions, transverse momenta and
virtuality of the first photon $k_1^2$ as follows:
 \bea
  \label{69}
   &&a_j\equiv -(p_1-k_j)^2+m^2\,,\;\;
     b_j\equiv  (p_3+k_j)^2-m^2\,,
   \nn\\ 
   &&a_1=a_1^0 +{1-x_1  \over x_1}\, k_1^2\,,\quad a_2=a_2^0\,,
   \nn\\
   &&b_1=b_1^0 +{x_1+X_3\over x_1}\, k_1^2\,,\quad b_2=b_2^0\,,
  \\
  &&a_{12}=a_{21}\equiv-(p_1-k_1-k_2)^2+m^2
  =a_{12}^0 +{X_3\over x_1}\,k_1^2\,,
  \nn\\
  &&b_{12}=b_{21}\equiv(p_3+k_1+k_2)^2-m^2 
  =b_{12}^0 +{1\over x_1}\,k_1^2 \,,
  \nn
 \eea
where we denote by the upper index $0$ the quantities in the limit
$k_1^2=0$ [given in (1.83)]. The virtuality $k_1^2$ depends on the
energy fractions $x_\pm$ with $x_++x_-=x_1$ and the transverse
momenta of the muons [see (\ref{virtk2})]
\be
  k_1^2= \frac{1}{x_+ x_-} \left[ x_1^2 m_\mu^2 +
  \left(x_- {\vec p}_{+\perp} - x_+ {\vec p}_{-\perp}\right)^2 \right]\,.
  \label{virtk12}
\ee

The contribution, corresponding to Fig.~\ref{fig:12},  can be
written in the form (we indicate explicitly the helicity states
$\Lambda_1$ ($\Lambda_2$) of the virtual (real) photon and the
initial and final electrons $\lambda_{1,3}$):
\bea
  &&J^{(\ref{fig:12})}_{1}= - {(4\pi\alpha)^{2}\over k_1^2}
  \label{70}\\
  &&
  \times
  \sum_{\Lambda_1=0,\pm 1}  \overline{V}^{\Lambda_1}_{\lambda_+
  \lambda_-}(k_1,p_+) \,C_{\lambda_1\, \lambda_3}^{\Lambda_1\,
  \Lambda_2}\, (x_1, x_2,
   k_{1\perp},  k_{2\perp}, p_{3\perp})\,,
  \nn
\eea
where $\overline{V}(k_1,p_+)$ is given by (\ref{25}) identifying
$k=k_1$ and $m=m_\mu$. To calculate $C$, we follow the electron
line from left to right in the diagrams of Fig.~\ref{fig:12} 
and write down
the corresponding vertices:  
\bea
  C &=& \frac{1}{a_1 a_{12}}\, V(p_1, k_1)\,
  V(p_1-k_1, k_2)\,V(p_3-q)
  \nn\\
  &-&
  \frac{1}{a_1 b_2}\, V(p_1, k_1)\,V(p_1-k_1)\, V(p_1-k_1+q, k_2)
  \nn\\
  &+&
  \frac{1}{b_{12} b_2}\,V(p_1)\, V(p_1 +q, k_1)\, V(p_1-k_1+q, k_2)
  \nn\\
  &-&
 \frac{1}{a_{12}}\,V(p_1, k_1,k_2)\, V(p_3-q)
  \nn\\
  &+&
  \frac{1}{b_{12}}\,V(p_1)\, V(p_1+q,k_1, k_2)
  \nn\\
  &+& (k_1 \leftrightarrow k_2)\,.
  \label{71}
  \eea
The last two contributions contain the four particle vertices
corresponding to two last diagrams of Fig.~\ref{fig:12} with the
crossed lines. Next we take into account the explicit expression
(15) for the vertices $V(p)$, the relations for $V(p,k)$ similar
to (1.86), the relation
$$
  V(p_1, k_1,k_2)=V(p_1+q,k_1, k_2)
$$
and present $C$ in the form
\be
   C=\sqrt{2}\,X_3 \left(1 +{\cal P}_{12}\right) \,
   M_{\lambda_1\, \lambda_3}^{\Lambda_1\, \Lambda_2}\, (x_1, x_2,
   k_{1\perp},  k_{2\perp}, p_{3\perp}) \, \Phi_{13}\,.
\ee
Here we have introduced the permutation operator ${\cal P}_{12}$ and
the factor
\be
  \Phi_{13}= \frac{1}{\sqrt{X_3}} \,
  {\rm e}^{{\rm i} (\lambda_3 \varphi_3 - \lambda_1 \varphi_1)}
  \label{73}
\ee
including the common phase. This allows us to omit below all
factors $\Phi$ from vertices $V(p,k)$ and $V(p,k_1,k_2)$. As a
result, we obtain the expression similar to (1.87), but with
$\Lambda_1=0,\pm 1$:
\bea
  X_3 M_{\lambda_1\,\lambda_3}^{\Lambda_1 \Lambda_2}&=&
  A_2 \, V_{\lambda_1\lambda}^{\Lambda_1}(p_1,k_1) \,
         V_{\lambda\lambda_3}^{\Lambda_2}(p_1-k_1 +q, k_2)
  \nn\\
  &+&
  q_\perp {B_2}_{\lambda_1\,\lambda_3}^{\Lambda_1 \Lambda_2}+
   {\widetilde A}_2 V_{\lambda_1\lambda_3}^{\Lambda_1 \Lambda_2}(p_1,\,
  k_1,\,k_2)\,,
    \label{74}
\eea
where the quantities $A_2$ and $\widetilde{A}_2$ are of the same
form as in (1.88), however with denominators $a_1$, $b_1$,
$a_{12}$ and $b_{12}$ of the propagators depending on the
virtuality $k_1^2$:
\be
  A_2=  \frac{X_3}{a_1a_{12}} - \frac{1-x_1}{a_1b_2}+
  \frac{1}{b_{12} b_2}\,,\;\;\;
  {\widetilde A}_2 =-\frac{X_3}{a_{12}} + \frac{1}{b_{12}}\,.
  \label{75}
 \ee
The transverse 4-vector $B_2$ is similar to that in (1.89):
\bea
  {B_2}_{\lambda_1\,\lambda_3}^{\Lambda_1 \Lambda_2} &=&
  - X_3\,{2 {e}^{(\Lambda_2)\,*}_\perp\over a_1a_{12}}\,
  V_{\lambda_1 \lambda_3}^{\Lambda_1}(p_1,k_1)
  \nn\\
  &&\times
  \left( 1- {x_2\over 1-x_1}\, \delta_{\Lambda_2,- 2\lambda_3}
  \right)
  \\
  &&+{2 {e}^{(\Lambda_1)\,*}_\perp\over  b_{12} b_2}\,
  V_{\lambda_1 \lambda_3}^{\Lambda_2}(p_1-k_1+q, k_2)
  \nn\\
  &&\times
  \left( 1- x_1\,
  \delta_{\Lambda_1,- 2\lambda_1} \right)\,
    \left(1-\delta_{\Lambda_1,0}\right)\,.
  \nn
  \label{76}
\eea

For the case $\Lambda_1=\pm 1$ all independent helicity states of
amplitude $M_{\lambda_1\,\lambda_3}^{\Lambda_1 \Lambda_2}$ from
(\ref{74}) coincide with those in (1.92)--(1.96). For the case
$\Lambda_1=0$ we obtain
\bea
  && X_3 M_{\lambda_1\,\lambda_3}^{0 \Lambda_2}=
  2\sqrt{2k_1^2}\,{1-x_1\over x_1}\Bigg[ A_2 \,
  V_{\lambda_1\lambda_3}^{\Lambda_2}(p_1-k_1,k_2)
  \nn\\
  &&~
  \label{77}\\
  &&
  - \frac{X_3}{a_1 a_{12}} 
  \left(q_\perp {e}^{(\Lambda_2)\,*}_\perp\right)
  \left( 1- {x_2\over 1-x_1}\, \delta_{\Lambda_2,- 2\lambda_3}
  \right)\delta_{\lambda_1,\lambda_3}\Bigg]\,.
  \nn
\eea
The amplitudes (\ref{74}), (\ref{77}) are given in such a form
that all individual large (compared to $ q_\perp$) contributions
have been rearranged into finite expressions. To show that the
impact factor $J_1^{(\ref{fig:12})}\propto q_\perp$, it is
sufficient to check that the quantities $A_2$ and ${\widetilde
A}_2$ vanish in the limit of small $q_\perp$:
\be
  A_2\propto q_\perp\,, \;\;
  {\widetilde A}_2 \propto q_\perp \,.
  \label{78}
\ee
This could  be done by a direct substitution of the expressions  for
the denominators (\ref{69}) into (\ref{75}).

However, it is much easier to be proved using the following simple
consideration. Since $x_q = 2qP_2/s \sim m^2/s$ and $y_q = 2qP_1/s
\sim m^2/s$, the quantity $q^2 = sx_qy_q+q^2_\perp \approx
q^2_\perp$ tends to zero in the limit $q_\perp \to 0$. Therefore,
in this limit we have
\bea
  a_{12}= 2p_3q-q^2 \to X_3 sy_q,&& b_{12}= 2p_1q+q^2 \to s
  y_q, 
  \nn\\
  a_{12}&\to& X_3 b_{12}\,.
\eea
Taking into account
\be
a_1=-(p_3+k_2-q)^2+m^2 \to -b_2 +(1-x_1) sy_q
\ee
this leads to
\bea
\widetilde{A}_2 &\to& 0\,,
\nn\\
A_2 &\to& \frac{1}{a_1 sy_q} -
\frac{1-x_1}{a_1b_2}+  \frac{1}{sy_q b_2}
\\ 
&\propto& b_2 - (1-x_1) sy_q + a_1 \to 0\,.
\nn
\eea

\section{Summary}
\label{summary}

In the present paper we continued to develop a new effective
method for calculating all helicity amplitudes of jet-like QED
processes at tree level.

Using the jet-like kinematics, the scattering amplitudes are
represented in the simple factorized form (\ref{2}), where the
impact factors $J_1$ or $J_2$ are proportional to the scattering
amplitudes of the first or second initial particle (lepton,
antilepton, photon) with the virtual $t$-channel photon
connecting the two impact factors. The final particles in those
two produced ``jets'' have emission and scattering angles much
less than unity, though they are allowed to be of the order of
typical emission angles $m_i/E_i$ or larger.

In calculating the impact factors in our kinematics we have
replaced the spinor structure involving leptons or antileptons of
small virtuality by transition vertices which are matrices with
respect to incoming and outgoing lepton helicities. These vertices
are finite in the limit $s\to \infty$.

In our previous paper we have considered multiple photon
bremsstrahlung. In that case the impact factors are given as
simple matrix products of vertices going along the lepton line.
One generic vertex describes the coupling of the leptons to the
$t$-channel virtual photon. At most two other nonzero such
vertices were needed to describe all processes with only real
bremsstrahlung photons.
At this stage the diagrams of Figs.~1.2, 1.4, 1.9 and 1.11
could be easily calculated including all helicity states.

In the present paper we extended our method to consider also
processes with lepton pair production. In other words, we now
allow that more than one lepton line connected by virtual photons
(with finite energy fraction and small virtuality) are present in
the considered impact factor.

The idea based on gauge invariance consists  in decomposing the
impact factor with such a virtual photon $k$ with helicity
$\Lambda$ into a product of two building blocks $A^{(\Lambda)}$
and $B^{(\Lambda)}$ (which contain their own lepton lines) and sum
over the helicities of the virtual photon:
 \be
J_1=  \frac{1}{k^2} A_\mu B^\mu= - \frac{1}{k^2}
\sum_{\Lambda=0,\pm1} A^{(\Lambda)} B^{(\Lambda)}. \label{blocks}
 \ee
The block $A^{(\Lambda)}= A e ^{(\Lambda)*}$ contains the
``outgoing'', $B^{(\Lambda)}= B e ^{(\Lambda)}$ --- the
``incoming'' virtual photon.

For that purpose we have generalized our previous bremsstrahlung
vertices to include the case of virtual photons, the corresponding
expressions are given in (\ref{16}). Using simple crossing
relations we found the vertices for the
$\gamma(k)\to e^+(p_+) + e^-(p_-)$ transition where the initial
photon is either real or virtual, the results are collected in
(\ref{25}) and (\ref{28}).
We have also introduced vertices with four external lines
(\ref{19}), (\ref{30})---(\ref{32}) (analogous to the case of
scalar QED) for the
 $e(p)\to [\gamma(k_1)\gamma(k_2)]+e(p')$ as
well as the $\gamma(k)\to [e^+(p_+) \gamma(k')] + e^-(p_-)$ and
$\gamma(k)\to [e^-(p_-)\gamma(k')] + e^+(p_+)$ transitions.
Using these vertices we develop the convenient diagrammatic rules
presented in Figs.~\ref{fig:3}---\ref{fig:7}.

To discuss the impact factor for the case where initial photons
and final leptons or initial leptons and final photons are
interchanged we have presented the corresponding crossing rules
in (\ref{43})--(\ref{45}). So, e.g., the impact factor of diagram
of Fig.~1.10 is the cross channel of the double bremsstrahlung
impact factor of Fig.~1.9.

Let us recall again, that the impact factors are finite in the
high-energy limit: they depend on the energy fractions and the
transverse momenta of the jet particles, and on the helicities of
all initial and final particles. By construction all individual
large contributions (compared to $q_\perp$) are arranged into
finite expressions. Therefore, these helicity amplitudes are very
convenient for numerical calculations in the jet-like kinematics.

We have applied our technique to calculate the impact factor for
the single lepton pair production of Fig.~1.3, see (\ref{50}).
This allows to obtain the pair production processes shown in
Figs.~1.3 and 1.5. Taking into account also the impact factors for
single bremsstrahlung, the diagram of Fig.~1.6 is captured too.

The impact factor for the pair production of Fig.~1.8 has been
studied in Sect.~\ref{sec:4}, see (\ref{56}), (\ref{61}),
(\ref{61a}). Using the crossing relations, also the reaction of
Fig.~1.7 is described.

Finally we demonstrated in Sect.~\ref{sec:5} how our new meth\-od
can be used to calculate the leading contribution to higher order
impact factors taking as an example a jet from an electron
containing in addition a muon pair and a photon. This impact
factor can be used to describe the process of Fig.~\ref{fig:1}.

\section*{Acknowledgements}

We are grateful to S.~Brodsky and A.~Vainshtein for  useful
discussions. This work is supported in part by INTAS (code
00-00679) and by RFBR (code 03-02-17734).

\end{document}